# IRON BEHAVING BADLY: INAPPROPRIATE IRON CHELATION AS A MAJOR CONTRIBUTOR TO THE AETIOLOGY OF VASCULAR AND OTHER PROGRESSIVE INFLAMMATORY AND DEGENERATIVE DISEASES

Douglas B. Kell

School of Chemistry and Manchester Interdisciplinary Biocentre, The University of Manchester, 131 Princess St, MANCHESTER M1 7DN, UK

dbk@manchester.ac.uk     www.dbkgroup.org     www.mcisb.ac.uk

# Background and preamble

This is an unusual experiment in the Open Access scientific publication of a longish pre-print. Its story is as follows. (The Table of Contents is on p3.)

While visiting Columbia University to present a seminar in May 2007, I met Jon Barrasch, who described, and got me interested in, his work on NGAL and kidney disease. We discussed some ideas for experiments. On my return, I started reading into this literature to do my homework, found it fascinating, started making notes, and eventually determined that:

- many of the features of kidney disease were replicated in a great many other diseases
- the literatures of these different diseases barely overlapped but there were clear themes, such as oxidative stress and inflammation, that were common to them
- when one looked for additional evidence of a contribution of poorly liganded iron to this oxidative stress there was in many case a <u>considerable</u> literature pointing up its involvement
- even when I thought I had trawled the literature exhaustively I continued (and still continue) to find highly pertinent papers, even papers in major journals that have been cited many times; this problem is exacerbated by the fact that most journals restrict the number of citations
- this 'balkanisation' of the literature is also in part due to the amount of it (some 25,000 journals with presently 2.5 million <u>peer-reviewed</u> papers per year, i.e. ~ 5 **per minute** [1]), with a number http://www.nlm.nih.gov/bsd/medline_cit_counts_yr_pub.html approaching 2 per minute at Medline
- the disconnect between the papers in the literature (usually as pdf files) and the metadata describing them (author, journal, year, pages, etc) is acute and badly needs filling [2]
- even though this preprint cites over 2000 literature sources, it still has many intellectual gaps
- when it became clear that these notes would have much greater value if written in the style of a review and then published I cohered them accordingly, and sent the result to three journals (including Open Access ones). They declined even to send the paper for review by external referees, and consequently this pre-print has not been peer-reviewed
- I shall be taking up a different job shortly that will mean that my scope for pursuing these issues is more limited than I had anticipated
- Consequently I have decided to make this work freely available at www.arxiv.org under (see http://creativecommons.org/licenses/by-sa/3.0/) the Creative Commons Attribution-Share Alike 3.0 Unported license, that permits <u>attributed</u> use and re-use. I thus request users to cite the arxiv.org URI when making use of the contents of this article, pending possible peer-reviewed publication elsewhere.

# Abstract

The production of peroxide and superoxide is an inevitable consequence of aerobic metabolism, and while these particular 'reactive oxygen species' (ROSs) can exhibit a number of biological effects, they are not of themselves excessively reactive and thus they are not especially damaging at physiological concentrations. However, their reactions with poorly liganded iron species can lead to the catalytic production of the very reactive and dangerous cellular hydroxyl radical, which is exceptionally damaging, and a major cause of chronic inflammation. We review the considerable and wide-ranging evidence for the involvement of this <u>combination</u> of (su)peroxide and poorly liganded iron in a large number of physiological and indeed pathological processes and inflammatory disorders, especially those involving the progressive degradation of cellular and organismal performance. These diseases share a great many similarities and thus might be considered to have a common cause (i.e. iron-catalysed free radical and especially hydroxyl radical generation). The studies reviewed include those focused on a series of cardiovascular, metabolic and neurological diseases, where iron can be found at the sites of plaques and lesions, as well as studies showing the significance of iron to aging and longevity. The effective chelation of iron by natural or synthetic ligands is thus of major physiological (and potentially therapeutic) importance. As systems properties, we need to recognise that physiological observables have multiple molecular causes, and studying them in isolation leads to inconsistent patterns of apparent causality when it is the simultaneous combination of multiple factors that is responsible. This explains, for instance, the decidedly mixed effects of antioxidants that have been observed, since in some circumstances (especially the presence of poorly liganded iron) molecules that are nominally antioxidants can actually act as pro-oxidants. The reduction of redox stress thus requires suitable levels of both antioxidants and effective iron chelators. Some polyphenolic antioxidants may serve both roles. Understanding the exact speciation and liganding of iron in all its states is thus crucial to separating its various pro- and anti-inflammatory activities. Redox stress, innate immunity and pro- (and some anti-)inflammatory cytokines are linked in particular via signalling pathways involving NF-κB and p38, with the oxidative roles of iron here seemingly involved upstream of the IκB kinase (IKK) reaction. In a number of cases it is possible to identify mechanisms by which ROSs and poorly liganded iron act synergistically and autocatalytically, leading to 'runaway' reactions that are hard to control unless one tackles multiple sites of action simultaneously. Some molecules such as statins and erythropoietin, not traditionally associated with anti-inflammatory activity, do indeed have 'pleiotropic' anti-inflammatory effects that may be of benefit here. Overall we argue, by synthesising a widely dispersed literature, that the role of poorly liganded iron has been rather underappreciated in the past, and that in combination with peroxide and superoxide its activity underpins the behaviour of a great many physiological processes that degrade over time. Understanding these requires an integrative, systems-level approach.

# Contents

## Introduction

Even under 'normal' conditions, as well as during ischaemia when tissue oxygenation levels are low, the redox poise of the mitochondrial respiratory chain is such that the normally complete four-electron reduction of dioxygen to water is also accompanied by the production, at considerable rates (ca 1-4% of $O_2$ reduced), of partially reduced forms of dioxygen such as hydrogen peroxide and superoxide (e.g. [3-17]). These 1- and 2-electron reductions of $O_2$ are necessarily exacerbated when the redox poise of the b-type cytochromes is low, as when substrate supplies are in excess or when the terminal electron acceptor $O_2$ is abnormally low due to hypoxia or ischaemia. Various other oxygenases, oxidases and peroxidases can also lead directly to the production of such 'reduced' forms of dioxygen *in vivo* (e.g. [18-20]), with $H_2O_2$ from xanthine oxidase being especially implicated in ischaemia/reperfusion injury (e.g. [19; 21-26]). These molecules can thus cause or contribute to various kinds of oxidative stress. However, this is mainly not in fact because they can react directly with tissue components themselves, since they are comparatively non-toxic, and are even used in cellular signalling (e.g. [27-30]), but much more importantly because they can react with other particular species to create far more reactive and damaging products such as hydroxyl radicals, with all these agents being collectively known as reactive oxygen species (ROSs). Possibly the commonest means by which such much more damaging species are created is by reaction with unliganded or incompletely liganded iron ions [31-33]. **The themes of this review are (i) that it is this combination of poorly liganded iron species, coupled to the natural production of ROSs, that is especially damaging, (ii) that the role of iron has received far less attention than has the concept of ROSs, albeit the large literature that we review, and (iii) that this combination underpins a great many (and often similar) physiological changes leading to disease manifestations, and in particular those where the development of the disease is manifestly progressive and degenerative.**

An overview of the structure of the review is given in Fig 1, in the form of a 'mind map' [34]. The literature review for this meta-analysis was completed on June 30th, 2008.

## Some relevant chemistry of iron and reduced forms of oxygen

While superoxide and peroxide are the proximate forms of incomplete $O_2$ reduction in biology, a reaction catalysed by the enzyme superoxide dismutase [35] serves to equilibrate superoxide and peroxide:

$$2\ O_2^{\bullet-} + 2H^+ \rightarrow H_2O_2 + O_2 \tag{1}$$

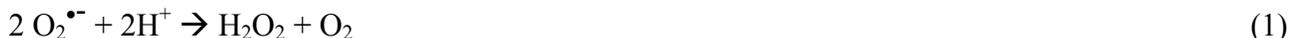

Arguably the most important reaction of hydrogen peroxide with (free or poorly liganded) Fe(II) is the Fenton reaction [36], leading to the damaging hydroxyl radical ($OH^{\bullet}$)

$$Fe(II) + H_2O_2 \rightarrow Fe(III) + OH^- + OH^{\bullet} \tag{2}$$

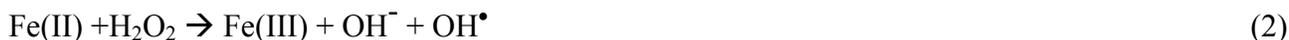

Superoxide can also react with ferric iron in the Haber-Weiss reaction [37] to produce Fe(II) again, thereby effecting redox cycling:

$$O_2^{\bullet-} + Fe(III) \rightarrow O_2 + Fe(II) \tag{3}$$

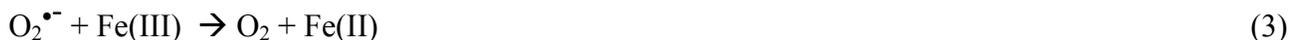

Ascorbate can replace $O_2^{\bullet-}$ within the cell for reducing the Fe(III) to Fe(II) [38]. Further reactions, that are not the real focus here, follow from the ability of hydroxyl radicals and indeed Fe(n) directly to interact with many biological macro- and small molecules, especially including DNA, proteins and unsaturated lipids. Thus [39-43], Fe(II) and certain Fe(II) chelates react with lipid hydroperoxides (ROOH), as they do with hydrogen peroxide, splitting the O—O bond. This gives $RO^{\bullet}$, an alkoxyl radical, which can also abstract $H^{\bullet}$ from polyunsaturated fatty acids and from hydroperoxides. The resulting peroxyl radicals $ROO^{\bullet}$ can continue propagation of lipid peroxidation. Oxidative stress also leads to considerable DNA damage [44-46] and to the polymerisation and denaturation of proteins [47-49] and proteolipids that can together form insoluble structures typically known as lipofucsin (see e.g. [50; 51]) or indeed plaques. Such plaques can also entrap the catalysts of their formation. These are described below. Many small molecule metabolic markers for this kind of **oxidative stress** induced by the hydroxyl radical and other 'reactive oxygen species' (ROSs) are known [16; 52-59], and include 8-oxo-guanine [60-64], 8-hydroxy guanine [65], 8-hydroxy-2'-deoxy-guanosine [66; 67], 8-oxo-GTP [68], 4-hydroxy-2-hexenal [69], 4-hydroxy-nonenal [70], 4-hydroperoxy-2-nonenal, various isoprostanes [71-77], 7-keto-cholesterol [78], malondialdehyde [79], neopterin [80], nitrotyrosine [81-84] and thymidine glycol [85; 86]. Note that the trivial names in common use for this kind of metabolite are not helpful and may even be ambiguous or misleading, and it is desirable (e.g. [87]) to refer to such molecules using terminology that relates them either to molecules identified in persistent curated datbases such as ChEBI [88] or KEGG [89], or better to their database-independent encodings as SMILES [90] or InChI [91-93] strings. (There are other oxidative markers that may be less direct, such as the ratio of 6-keto-prostaglandin F1α to thromboxane B2 [94], but these are not our focus here.) Overall, it is in fact well established that the interactions between 'iron' *sensu lato* and partly reduced forms of oxygen can lead to the production of the very damaging hydroxyl radical (e.g. [16; 95-104]), and that this radical in particular probably underpins or mediates many higher-level manifestations of tissue damage, disease, organ failure and ultimately death [10; 102; 105-107]. While the role of ROSs in these processes has been widely discussed, the general recognition of the importance of inadequately liganded iron in each of them has perhaps been less than fully appreciated. One of our tasks here will therefore be to stress this role of 'iron', and to assess the various means of chelating 'iron' such that it does not in fact do this. (Throughout we use 'iron' to refer to forms of Fe(n, n>0) with unspecified ligands.)

For completeness we note the reactions catalysed by superoxide dismutase

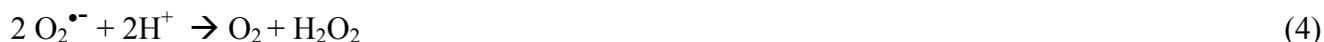

$$2\ O_2^{\bullet-} + 2H^+ \rightarrow O_2 + H_2O_2 \tag{4}$$

and by catalase

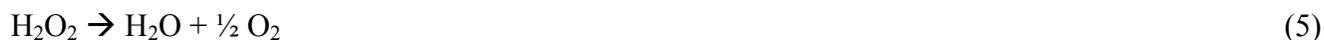

$$H_2O_2 \rightarrow H_2O + \tfrac{1}{2}\ O_2 \tag{5}$$

That together, were their activity in the relevant locations sufficiently great, might together serve to remove (su)peroxide from cells completely.

In addition to reactive oxygen species there are ions such as the perferryl ion (Fe-O) [108] and reactive nitrogen species [109; 110]. These latter are mainly formed from the natural radical NO, an important inflammatory mediator [111], with peroxynitrite production (from the reaction of NO and superoxide) [18; 112-115] leading to nitrotyrosine [81], or nitro-fatty acids formation [116; 117] being a common means of their detection downstream. Other toxic products of the reactions of NO include $NO_2$, $N_2O_3$, and S-nitrosothiols [118], and the sequelae of some of these may also involve iron [119].

Overall, we recognise that these kinds of inflammatory, oxidative stress-related reactions are accumulative and somewhat irreversible, that they are consequently age-related, and (see [120-123] and later) that most diseases and causes of mortality that are prevalent in the developed world are in this sense largely manifestations of this kind of aging.

## Ligands and siderophores

As well as the reactions described above, ferrous ions will react with oxygen under aerobic conditions to produce ferric ions, and in natural environments there is little to stop this. Consequently, and because these reflect fundamental physicochemical properties of such ions, the problems of both solubility and toxicity were faced by bacteria (and indeed fungi [124; 125]) long ago in evolution, and were solved by their creation and excretion of (mainly ferric-)iron chelators known as siderophores [126-145] (and for haemophores see [146]). These typically have extremely tight binding constants ($K_f$ > $10^{30}$ [147]) and can solubilise and sequester iron such that it can be internalised via suitable transporter molecules within the bacterial plasma membrane [148]. Bacterial and fungal siderophores usually form hexadentate octahedral complexes with ferric iron and typically employ hydroxamates, $\alpha$-hydroxycarboxylates and catechols as extremely effective $Fe^{3+}$ ligands [138]. Since bacterial growth requires iron, it is unsurprising that siderophores are effectively virulence factors (e.g. [130; 149-152]). While upwards of 500 microbial siderophores have been identified [138], with new ones still appearing [153], and with the most common one in medical use, desferrioxamine or DFO, being such a bacterial product (see below), it is an astonishing fact that no human siderophore has been chemically identified, even though such activities were detected nearly 30 years ago [154; 155] (see also [156-161]). As noted by Kaplan [162], "a discovery that mammals produce siderophores would lead to an epochal change in the paradigm of mammalian iron homeostasis." To this end, some recent events have begun to change matters, and our overall knowledge of the regulation of iron metabolism, considerably.

## Mammalian iron metabolism

The total body iron in an adult male is 3000 to 4000 mg and the daily iron requirement for erythropoiesis, the major 'sink', is about 20 mg [163]. However, the loss of iron in a typical adult male is very small [164] and can be met by absorbing just 1 - 2 mg of iron per day [165; 166]. This careful conservation and recycling of iron from degrading erythrocytes is in fact essential because typical human diets contain just enough iron to replace the small losses, although when dietary iron is more abundant, absorption must be (and is) attenuated since higher levels than necessary lead to iron overload and many distressing sequelae contingent on the radical production described above.

A variety of aspects of mammalian iron metabolism have been reviewed in detail elsewhere (e.g. [99; 104; 151; 167-186]), including a series on 'iron imports' [187-193], and for our present purposes (Fig 2) mainly involves the intestinal (mainly duodenal) uptake of Fe(II) (produced from Fe(III) using a luminal ferrireductase) via a divalent metal ion transporter DMT1/DCT1/NRAMP [194; 195] and its subsequent binding as Fe(III) to transferrin (Tf). The intestinal uptake of haem (heme) occurs via the heme carrier protein-1 (HCP1) [196] and internalized, while the iron in heme is liberated by heme oxygenase-1 (HO1) [197-199]. Haem is syntheses in many tissues, especially liver and erythroid cells [200]. Vesicular routes of intestinal transfer may also occur [201; 202]. Low MW cytoplasmic chelators such as citrate can bind iron fairly weakly and thereby contribute to a labile iron pool (LIP) in the cytoplasm and especially the lysosomes and mitochondria (see [203-206]), while ferritin too can bind cytoplasmic iron (via a chaperone [207]) and is seen as a good overall marker of iron status [208]. Iron(II) is subsequently exported through the basolateral membrane of the enterocyte by ferroportin-1 (FPN1) [209-211]. Ferroportin may also contribute to uptake in enterocytes [212]. Fe(III) may then be produced by hephaestin (Hp) [213] before it is bound by transferrin (Tf), which is the main but not sole means of binding Fe(III) when it is transported through the circulation, with major iron storage taking place in the liver. Similar processes occur in the peripheral tissues, with significant transfer of iron from transferrin occurring via the transferrin receptor [214].

'Free' haem appears in the circulation (it may have a signalling role [215]) and elsewhere largely because of erythrocyte degradation, and it can also greatly amplify the cellular damage caused by ROSs [216], and its degradation pathway via haem oxygenase [217; 218] to biliverdin and then using biliverdin reductase to form bilirubin generates 'free' (and potentially redox-active) iron. It would appear that, not least because biliverdin has powerful antioxidant properties, haem oxygenase is more protective than damaging [198; 219-223], even though one of the products of its reaction is Fe that must eventually be liganded (or e.g. incorporated into ferritin). (Another product is the gas CO, that has been proposed as a measure of oxidative stress in the lung [224].)

All of the above obviously ignores both some important aspects of the speciation and ligunding of iron, as well as the tissue distribution of the specific proteins involved – for which latter global information will shortly emerge [225] (www.proteinatlas.org/). It also ignores any discussion on the genetic regulation of iron metabolism (e.g. [226-229]), which is not our main focus.

However, since hepcidin has recently emerged as a 'master regulator' of regulation at the physiological level, we describe some of these new developments.

## *Hepcidin*

In the liver and elsewhere, many aspects of iron metabolism are regulated by a recently discovered 25-amino acid polypeptide called hepcidin [163; 186; 190; 212; 230-255] that acts in part as a negative regulator of iron efflux [256] by causing the internalisation of ferroportin [257-261]. Hepcidin is produced, partly under the regulation of a receptor called hemojuvelin (e.g. [262]), via an 84-aa precursor called pre-pro-hepcidin and a 60mer called pro-hepcidin [245; 263; 264] although the active agent is considered to be the 25mer referred to above, and with the inactive precursors appearing not to be useful markers [265; 266].

Strikingly, anaemia and anoxia both suppress hepcidin production [190; 267; 268] (Fig 3), such that just while superoxide production is being enhanced by the anoxia there is more iron being absorbed from the intestine and effluxed into the circulation. In view of the inter-reactivity of superoxide and iron this could be anticipated to enhance free radical formation, leading to a positive feedback loop in which the problems are amplified: ischaemia/anoxia changes Fe(n) distribution leading to differential reactivity with the products of anoxia and thus further free radical production. However, hepcidin is overexpressed in inflammatory disease and is an early inflammatory marker [190; 269-271]. Its expression is positively controlled inter alia by SMAD4, and loss of hepatic SMAD4 is thus associated with dramatically decreased expression of hepcidin in liver and increased duodenal expression of a variety of genes involved in intestinal iron absorption, including Dcytb, DMT1 and ferroportin, leading to iron overload [272]. STAT3 is another positive effector of hepcidin expression [273; 274], and ROSs inhibit this effect [275], thereby creating a link between ROSs and Fe metabolism. To understand the exact roles of hepcidin in iron metabolism, it is going to be especially important to understand where it is expressed; fortunately, such studies are beginning to emerge [276].

Overall there is a complex interplay between positive and negative regulation and the organismal distribution of iron caused by changes in hepcidin concentration [277], with in many cases the hypoxic response (decreased hepcidin) seeming to dominate that due to inflammation (increased hepcidin) even when iron levels are high [278; 279]. Specifically, lowered hepcidin causes hyperferraemia. Hepcidin is also activated by p53 [280], and may play a role in the degradation of atherosclerotic plaques [281]. Another recently discovered protein that is crucially involved in human iron metabolism is NGAL or siderocalin, and while there is some evidence for their co-regulation [282], they have normally been studied separately.

## *NGAL (also known as lipocalin-2 or siderocalin)*

Lipocalins [283] are a diverse group of ligand-binding proteins that share a conserved structure even in the absence of significant sequence conservation. This core structure includes an eight-stranded anti-parallel β barrel that defines a calyx, or cup-shaped structure, enclosing the ligand binding site.

NGAL – neutrophil gelatinase-associated lipocalin – is a 21kDal glycoprotein first isolated by Kjeldsen and colleagues in 1993 [284]. Synonyms include lipocalin 2, siderocalin, Lcn2, α2-microglobulin-related protein or neu-related lipocalin (in rats) [285; 286] and (in mice) 24p3 or uterocalin [287]. Although lipocalins are well known to be involved in the sequestration and transport of a variety of ligands, the natural ligand of NGAL (as is the case with many lipocalins) was not initially known even in terms of its chemical class. This changed with the seminal paper of Goetz and colleagues [288] (and see [162]) who purified recombinant NGAL from *E. coli* and found that its structure contained a negatively charged ferric siderophore with a subnanomolar dissociation constant that it had extracted from its bacterial host, and that the apo form of this molecule could also act as a potent bacteriostatic agent by sequestering iron (see also [289-293]). A companion paper [294] showed that iron-delivering activity was expressed in mammalian cells. The structure of NGAL is now known [295] and one of its interaction partners is a matrix metalloproteinase [296] to which it can presumably donate a metal ion and in the complex decrease its degradation [297].

The finding that NGAL was one of the most highly expressed proteins following ischaemia-reperfusion injury in kidney cells [298-300], and prognostic of kidney damage long before the more traditional marker creatinine was raised significantly, has led to considerable interest in this protein, especially as a marker of renal injury [301-311], and perhaps as a therapeutic [301]. Devireddy and colleagues [312] identified a receptor that internalizes 24p3, and internalization of iron bound to 24p3 prevents apoptosis. In contrast, internalization of the apo form of 24p3 that does not contain iron led to cellular iron efflux and apoptosis via the proapoptotic protein Bim [313]. In humans the megalin receptor can bind siderocalin (and its siderophore payload) and mediate its intracellular uptake [314]. Oxidative stress can also induce its expression [315].

Exogenously administered NGAL also markedly upregulates heme oxygenase-1, a proven multifunctional protective agent in experimental Acute Kidney Injury (AKI) that is thought to work by limiting iron uptake, promoting intracellular iron release, enhancing production of antioxidants such as biliverdin and carbon monoxide, and inducing the cell cycle regulatory protein p21 [220; 316; 317]. Because of this multifaceted protective action, NGAL has emerged as a prime therapeutic target in ischemic AKI [305].

Structural and direct binding studies have suggested that siderocalin tends (although not exclusively) to bind catecholate-type ligands, rather than hydroxamate- or carboxylate-based siderophores, at least when tested with microbially derived siderophores [288; 289; 291] (but cf. [295] for claims, disputed [286], as to the binding of bacterially derived formyl peptides!). The role of NGAL, as a siderophore-binding agent, is thus consistent with the widespread recognition that iron-induced radical generation is intimately involved in a variety of renal and other diseases [318; 319]. However, while it is certainly the case that siderocalin can reduce the virulence of bacteria when it binds the relevant bacterial siderophores [288-293] and that bacteria can 'evade' this by synthesising siderophores that siderocalin cannot bind (e.g. [142; 143; 320; 321]), it is questionable whether the only role of siderocalin lies in fact in its antibacterial activity. Rather we would suggest that its main role is in sequestrating iron via a human siderophore to stop inappropriately liganded iron from producing damaging oxygen radicals. Consistent with the liganding role is the fact that the tissue most highly expressing NGAL under normal conditions is bone marrow [286; 322], the site of erythropoiesis. The liganding can be extensive; as Goetz and colleagues [288] note, "During inflammation, concentrations of NGAL can increase to levels, with concentrations approaching 20–30 nM in the serum [323], presumably adequate to bind all available iron as ferric siderophore complexes".

Significant changes in NGAL expression have also been observed, for instance, during kinase-mediated signalling [324; 325], in cardiovascular disease [326-329] and in cancer [330; 331].

These findings on the kidney and the role of NGAL, together with the important knowledge that its chief ligand is probably an unknown human siderophore (Figs 2,4), thus lead us to consider the role of this system (and unliganded iron generally) in a whole series of other diseases that all share many characteristics of oxidative stress and inflammation (see also [332]). A similar thesis, albeit with comparatively little stress on iron, is the *leitmotif* of Finch's recent detailed monograph [121]. The theme of these sections is thus to stress the fact that while the role of ROSs in such syndromes has been pointed up previously, that of iron has not so generally yet been stressed, notwithstanding that there is in fact a great deal of pertinent literature that we here highlight as the focus of this review.

# Some disease manifestations in which iron may be implicated

## *Preeclampsia (PE)*

Another important disease that shares many of the same properties (or at least sequelae) of renal impairment, and may have the same fundamental aetiology, is pre-eclampsia. This is the most significant cause of morbidity and mortality in pregnant women [333]. The chief clinical manifestations at time of diagnosis are a raised blood pressure (hypertension) [334] and proteinuria, together with raised creatinine, consistent with the underlined reversible existence (since it is relieved upon delivery of the baby) of renal impairment. However, prognostic markers that might manifest early in pregnancy are lacking, and would be highly desirable. There is widespread agreement [335] that a poor connection of the placenta to the uterus leads to ischaemia and thus oxidative stress, with a substantial involvement of apoptosis during the placental remodelling [336-342]. Since preeclampsia-like syndromes can be induced in pregnant animals by surgical restriction of the uteroplacental blood supply [343], it is presumed that blood-borne agents arising from the ischemic placenta are the cause of the generalized endothelial cell damage and inflammatory responses that give rise to the symptoms of hypertension, proteinuria, and sudden oedema characteristic of preeclampsia [40]. Indeed, many studies implicate oxidative stress as a substantial contributor to this [344-385] [386-408], while some have noted the importance of iron status [40; 98; 369; 409-427], and so far as is known the transporters of iron in the placenta are similar to those in other cells [428]. Oxidative stress of this type is of course inflammatory in nature and inflammation is observed in PE [391; 395; 403; 405; 429-435]. We suggest strongly that it is the underline{combination} of inadequately liganded Fe(II or III) and superoxide/peroxide leading to OH$^\bullet$ formation that is the chief mechanistic cause of the downstream events that manifest in PE, and that appropriate removal by liganding/chelation or otherwise of these ions would prove of therapeutic benefit. (Iron status has also been implicated in other pregnancy and neonatal disorders [436-440].) There is evidence too for the involvement of the radical NO [375; 441].

We note that it is quite common nevertheless for iron to be prescribed, during pregnancy, especially during its latter stages [442; 443], and that this does of course lead to oxidative stress [444; 445].

Oxidative stress is caused both by the initial rate of production of superoxide and the rate of their conversion into OH$^\bullet$ radicals. The former can be induced by hypoxic conditions such as occur at high altitude, and one prediction, that is borne out [406; 446], is that PE should therefore be more prevalent at high altitude. Erythropoietin may be a marker for oxidative stress in pre-eclampsia [447].

Regarding the second stage, predictions include that PE should be more common in those suffering from diseases of iron metabolism. Although such mothers are of course less well *a priori*, this prediction is borne out for α-thalassemia [448; 449] although not, interestingly, for haemochromatosis [450]. We note in this context that thalassaemia not only predisposes towards PE but is known in general to cause hepcidin to decrease and NGAL to increase [278; 279; 282; 451], with consequent and inevitable iron dysregulation.

Another prediction is then that hepcidin should be changed in pre-eclampsia. Although no serum measurements have been reported to date, it is of extreme interest that – while they took it to be an antimicrobial peptide rather than an iron regulator – a recent study by Knox and colleagues of placental gene expression in a mouse model of PE showed that hepcidin expression underline{increased} by a greater factor than that of any other gene save one [452], consistent with the view that major changes in the regulation of iron liganding and metabolism underpin PE.

Finally, we note that NGAL is significantly implicated in pregnancy, and was even named uterocalin in mice to reflect its high expression in the uterus [287; 453-455].

## *Diabetes*

Type 2 diabetes and insulin resistance are known complications of pregnancy (e.g. [456-460]), and also predispose towards PE. In a similar vein, various types of pregnancy-related intrauterine growth restriction predispose towards diabetes in later life [461; 462], pointing up the progressive nature of these syndromes. Metabolic biomarkers for the one can thus be predictive of the other [463], consistent with a common cause. Certainly ROSs are known to play a substantive role in both insulin resistance [464-471] and in a variety of diabetic sequelae [65; 472-474], and mitochondrial dysfunction may be an early step in this [475]. Some anti-diabetic drugs, such as the 'glitazones' that are considered to act on Peroxisome Proliferator Activated Receptor (PPAR)γ, may also act by decreasing ROS production (e.g. [476-480]), and are even active against cerebral ischaemia and stroke [481-484]. As with most if not all of the other diseases we review here, studies of pro-inflammatory markers (such as TNF-$\alpha$, IL-1 and C-reactive protein [485]) during the development of diabetes show its aetiology to be inflammatory in nature [468; 486-505]. Iron 'excess' is also a known feature of gestational diabetes [506-508], a clear risk factor for the disease even in 'normal' populations [509-514], and indeed diabetes is a classical consequence of iron overloading as seen in hereditary haemochromatosis [515]. Serum ferritin is strongly associated with diabetes [516-519], including prospectively [520], while changes in visfatin are also intimately involved in changes of iron metabolism (with pro-hepcidin being elevated) [521]. Most importantly, lipocalin 2 (siderocalin/NGAL) is strongly associated with the development of diabetes [522; 523]. Lowering iron improves insulin sensitivity [511; 524], and metallothionein is protective [525-529]. There seems little doubt that iron status is a major determinant of the development of type 2 diabetes [530].

Non-transferrin-bound iron is also considerably elevated in type 2 diabetes [531], and this too is exacerbated by vitamin C. Iron metabolism is substantially deranged in type 2 diabetes and the metabolism of glucose (a reducing sugar) interacts significantly with iron metabolism [511]. Iron is also strongly implicated in non-alcoholic steatohepatitis, considered an early marker of insulin resistance [532-534]. Well-known diabetic complications include retinopathies, and it is noteworthy that elevated levels of ferritin can lead to cataract formation [535; 536].

## The metabolic syndrome

Although some of its origins may be pre-natal [462], many of the features of these diseases are also seen in the Metabolic Syndrome [537-541]. Thus, serum ferritin is also related to insulin resistance [517; 542; 543] and iron levels are raised [534; 544]. Of course diabetes and the Metabolic Syndrome are also closely coupled, so it is reasonable that features observed in the one may be observed during the development of the other. The metabolic syndrome is also an independent indicator for chronic kidney disease [545] and may be related to liver steatosis [546]. Metabolic disorders of this type too are closely intertwined with inflammation [490; 496; 502; 547], that is of course stimulated by ROSs whose generation is increased by high-fat diets [548]. Thus, our role here is to point up the existence of a considerable body of more-than-circumstantial evidence that here too the progressive and damaging nature of these diseases may be caused, in part, by inappropriately chelated iron.

## Obesity

"As previously pointed out by Booth *et al.* [549], 100% of the increase in the prevalence of Type 2 diabetes and obesity in the United States during the latter half of the 20th century must be attributed to a changing environment interacting with genes, because 0% of the human genome has changed during this time period." [539]

It is well known that there has been a staggering increase in the prevalence of obesity, diabetes, and especially type 2 diabetes, in the last 50 years or so, and that this increase is expected to continue (e.g. [550-552] and http://www.who.int/diabetes/). Equally, it is now well known that obesity, metabolic syndrome, diabetes and cardiovascular diseases are all more or less related to each other [552], and the question arises here as to whether dysfunctional iron metabolism might be a feature of each of them. In the case of obesity *per se*, however, we see no major evidence as yet for a causative role of deranged iron metabolism or chelation in causing obesity. Indeed, what little evidence there is [553; 554] suggests that the converse may be true, i.e. that changes in iron metabolism might be consequent upon obesity (possibly via peroxide generation [548]). Importantly, considerable evidence suggests that obesity and inflammation are significantly related [121; 405; 490; 496; 551; 555-572], not least because adipocytes produce and release various adipokines including pro-inflammatory cytokines such as IL-6 and TNF-α [490; 557; 558; 573-579]. It is likely that it is the <u>combination</u> of overfeeding-induced obesity and inflammation (partly induced by the obesity itself [580]) that leads to diabetes [581]. Certainly there is evidence for increased ROS production in obese mice, possibly mediated in part via the fatty acid-induced activation of NAPH oxidase [582], while obesity is linked [583; 584] to urinary levels of 8-epi-PGF$_{2\alpha}$, a well established marker of oxidative stress (qv). Fig 5 summarises the above in a manner that stresses the roles of iron, overfeeding and inflammation in the genesis of these processes, and notes that interference in several of these steps is likely to be required to limit their progression to best advantage.

## *Hypertension*

As well its significance in pre-eclampsia (see above), hypertension is a well known risk factor for many cardiovascular and related disease (e.g. [585]), and there is considerable evidence that its underlying cause is inflammatory in nature [586-594], related to the metabolic syndrome and obesity (e.g. [556; 595-599]) and may be mediated mainly via ROSs [600]. Some of its sequelae may be mediated via iron [601; 602].

## *Cardiovascular diseases*

It is well known that elevated iron stores can predispose to coronary artery disease and thence myocardial infarction. The 'iron hypothesis' of the benefits of some iron depletion due to menstruation was devised to account for the lowering of heart-disease risk in young women (that disappears in those post-menopause) and was proposed by Jerome Sullivan in 1981 [603-605] (and see also [606; 607]). (In this sense, the lack of menstruation during pregnancy would predispose to a comparative abundance of iron, as is indeed found – see above.) It is of particular interest that the well-known adverse vascular effects of homocysteine (in inhibiting flow-mediated dilatation) are in fact iron-dependent [608-610], and that reducing homocysteine (e.g. by folate supplementation) <u>in the absence of lowering iron</u> has shown no clinical benefit to date [611], thereby suggestion iron mediation. By contrast, iron stores represent an established risk factor for cardiovascular disease [612].

Of course many factors such as lipid levels, stress, smoking and so are well-known risk factors for cardiovascular, coronary artery disease and related diseases. Indeed kidney disease is well established as a risk factor for cardiovascular disease [613-615] (and indeed stroke [616], all consistent with their having in part a common cause – we believe inflammation). Our purpose here, within the spirit of this review, is to indicate the evidence for the involvement of inappropriately chelated iron in cardiovascular diseases too. There is no doubt that the iron-mediated causal chain of ischaemia → (su)peroxide → OH$^\bullet$ radical formation occurs during the development of heart disease, especially during reperfusion injury [617-620], and suitable iron chelators inhibit this [621; 622] (see also [623; 624]). Iron is also involved in the protection that can be produced by ischaemic preconditioning [625; 626]. Erythropoietin, a hormone with multiple effects that may involve iron metabolism, is also protective [627; 628].

## Heart failure

The sequelae of heart failure are complex, and involve a <u>chronic and continuing</u> worsening of a variety of physiological properties. ROSs are certainly involved here, since allopurinol (a potent inhibitor of xanthine oxidase) improves prognosis considerably [629], and uric acid is a well known biomarker for heart failure (see e.g. [630; 631]). Biopyrrins, degradation products of bilirubin and thus markers of oxidative stress are also considerably increased [632]. Anaemia is a common occurrence (and risk factor) in heart failure [633-635], again implying a role for dysregulated iron metabolism and a need to understand the exact speciation of iron in chronic anaemias linked to inflammatory diseases [636].

It is next on the formation of atherosclerotic plaques that our attention is here focussed.

## Atherosclerosis

Atherosclerosis is a progressive <u>inflammatory</u> disease [637-667] characterized by the accumulation of both oxidised lipids and various fibrous elements in arteries, often as plaques [668; 669]. Iron and oxidised lipids are both found in atherosclerotic lesions [106; 670-677], and iron depletion by dietary or other means delays this [678-681]. There is a correlation between iron status and atherosclerosis [671; 682-691], plausibly caused in part by the known ability of poorly liganded iron to effect lipid [683] and protein peroxidation, and by the effects of primed neutrophils [692] and transferrin [664]. In this context, exogenous ferric iron is deleterious to endothelial function [693], while iron chelation improves it [694-697]. However, phlebotomy provided no clinical benefit here [698]. Note that iron levels in plaques correlate with the amount of oxidised proteins therein [675], and that in one study [672], the EPR-detectable iron (essentially Fe(III)) in atherosclerotic tissue was <u>seventeen times greater</u> than that in the equivalent healthy tissue; this is not a small effect.

Statins, typically developed on the basis of their ability to inhibit the enzyme HMG-CoA redutase and thus decrease serum cholesterol, are well established to have benefits in terms of decreasing the adverse events of various types of cardiovascular disease [699], albeit that in many populations (e.g. [700]) cholesterol is a poor predictor of cardiovascular disease. However, a known target of statins different from HMGCoA reductase is the $\beta_2$-integrin leukocyte function antigen-1 (LFA-1) [701; 702] and in this context, it is important to note that the clinical benefits of the statins are certainly <u>not</u> due solely to their cholesterol-lowering ability via the inhibition of HMG-CoA reductase (see e.g. [121; 222; 558; 701; 703-733]), and different statins can cause a variety of distinct expression profiles [734] that are inconsistent with a unitary mode of action. The apparent paradox [735] that lipid-lowering statins do indeed exhibit epidemiological disease-lowering benefit, while having little effect on plaques, is arguably well explained, especially within the context of the present review, via their additional anti-inflammatory effects [710; 712; 736-756] [121; 716-718; 723; 729; 730; 757-785], acting upstream of the nuclear transcription factor NF-κB (and see later). It is also extremely relevant, for instance, that some statins have metal chelating properties [786].

It has been pointed out that many measures of iron stress are inappropriate, since it is only the redox-active form of iron that is likely involved in oxidative stress. Serum ferritin is considered by some to be the most reliable marker of iron status in general [787], although it is not well correlated with iron distribution in the heart [788]. What is clear, however, from the above is that deranged iron metabolism is intimately and causally involved in the formation of atherosclerotic lesions, and that appropriate iron chelation can help both to prevent and to reverse this.

Iron status is also closely involved in other chronic vascular diseases, and in the behaviour of wounds [789-792].

## Stroke

Stroke is caused by ischaemia, leading to inflammation [793] and ROS and other damaging free radical formation [794], in the brain (which is high in metal ions [795]), and is exacerbated by existing inflammation – see e.g. [796; 797]. Thus, another prediction is that iron excess should also aggravate the sequelae of stroke, and that appropriate chelation or free radical trapping agents should mitigate these effects. These predictions are indeed borne out [103; 798-807]. It is also of considerable interest that plasma NGAL levels are increased in stroke [326]; it is noteworthy that this can be seen in plasma despite the localised origin of the disease.

A variety of other studies have shown the beneficial treatments in stroke models of anti-inflammatory and antioxidant treatment, i.e. treatments that lower the amount of ROSs (e.g. [808-815]), as well as of preconditioning [816]. Given its role in iron metabolism, it is of considerable interest that erythropoietin also seems to be very effective in protecting against brain ischaemia/reperfusion injury and stroke [817-831], by a mechanism independent of erythropoiesis [832-834], and one that appears to involve anti-inflammatory activity [835].

## Alzheimer's, Parkinson's and other major neurodegenerative diseases

Oxidative stress and inflammation are early events of neurodegenerative diseases [813; 836-849] such as Alzheimer's disease (e.g. [850-865]), where plaque formation precedes neurodegeneration [866]. Iron (and in some cases copper) is also strongly implicated in a variety of neurodegenerative diseases [867] [868] [869] [870] [871] [872] [836] [873] [837] [850] [874] [875] [876] [877] [854] [878] [879] [880] [881] [842] [882] [883] [884] [885] [886] [887] [888] [889] [890] [891] [892] [893] [894] [895] [896] [897] [898] [899] [900] [901] [902] [903] [904] [905] [906] [907] [908] [909] [910] [911] [912] [810] [913] [914] [915] [916] [917] [918] [919] [920] [921] [922] [923] [924] [925] [219] [926] [927] [928] [929] [930] [931] [932] [933] [934] [935] [936] [937] [938] [939] [940] [941] [942] [943] [944] [945] [16] [946] [947] [948] [949] [226] [950] [864] [951] [952] [953] [954] [955] [106] [956] [957] [958] [959] [960] [961] [962] [963] [964] [965] [966] [967] [968] [969] [970] [971] [972] [973].

Indeed Thompson and colleagues comment [101] that "The underlying pathogenic event in oxidative stress is cellular iron mismanagement" and stress that "Multiple lines of evidence implicate redox-active transition metals, such as iron and copper, as mediators of oxidative stress and ROS production in neurodegenerative diseases". There is also ample evidence for its presence in the plaques characteristic of Alzheimer's disease [895; 904; 918; 974], just as in those of atherosclerosis (see above). Note too that iron can catalyse the oxidation of dopamine to a quinine form that can bind covalently to and then aggregate proteins [975]. Kostoff [976] has used a very interesting literature-based discovery approach to highlight the role of oxidative stress in the development of Parkinson's disease.

Other papers highlight the role of iron in multiple sclerosis [792; 883; 977-984] and in prion diseases [859; 985; 986]. However, a particularly clear example of iron-mediated neurodegeneration is given by the sequelae consequent upon lesions in a protein known as frataxin.

## Friedreich's ataxia

A number of respiratory chain components contain non-heme iron, and the question arises as to how they acquire it. Frataxin is a mitochondrial iron chaperone protein [987-993], involved in the safe insertion of Fe(II) during the production of Fe-S centres in the mitochondrial respiratory chain [994]. As are some other aspects of iron metabolism [995] , it is highly conserved in eukaryotes from yeast to humans [996; 997], a fact that made the unravelling of its function considerably easier [998-1007]. Friedreich's ataxia (FA) is a neurodegenerative disorder that arises from a genetic deficit of frataxin activity, whether by a missense mutation or, much more commonly, via the addition of GAA trinucleotide repeats [991; 1008-1011]. As well as the neurodegeneration and measurable iron deposition, clinical symptoms include cardiac hypertrophy [1012] and (pre-)diabetes [1013], consistent with the general thesis described here that all are in part manifestations of iron dysregulation, and in which suitable chelation may be beneficial [1014; 1015] (but cf. [1016]).

ROSs are undoubtedly involved in FA [1017; 1018], specifically via Fenton chemistry [1019; 1020], since the attenuation of $H_2O_2$ production (but not of superoxide) [994] ameliorates the disease [1021]. The deficit in frataxin causes both an increase in ROS ($H_2O_2$) production via the mitochondrial electron transport deficiency [1022] as well as a dysregulation in iron metabolism, potentially a very damaging synergistic combination (see later). Elements of its (in)activity that are seen as paradoxical [1023] are in fact easily explained when one recognises that it is the underlined combination of free iron with $H_2O_2$ that is especially damaging.

## Amyotrophic lateral sclerosis (ALS) or Lou Gehrig's disease

ALS is another progressive inflammatory [1024]  disease in which motor neuron death causes irreversible wasting of skeletal muscles. It has largely defied efforts to uncover the genetic basis of any predisposition [1025], save for a very clear association with defects in a Cu/Zn superoxide dismutase [8; 1026-1028] that can obviously lead to an increase in the steady-state levels of superoxide (and hence hydroxyl radical formation). There is also significant evidence for the involvement of iron [1029; 1030]. Drug therapies have to date shown rather limited benefits, and more in mouse models of Cu/Zn SOD deficiencies than in humans, though iron chelation therapy does not seem to have figured heavily, and it is recognised that combination therapies might offer better prognoses [1031].

## Aging

Aging or senescence is defined as a decline in performance and fitness with advancing age [1032]. Iron stores tend to increase with age, partly due to dietary reasons [1033] (and see [1034]). So too does the expression of NGAL/Lcn2/siderocalin, a process that can be reversed by melatonin [1035]. Mainstream theories of aging [121; 123; 1036-1053] recognise the relationship between progressive inflammation, cellular damage and repair and the higher-level manifestations of the aging process, and ('the free radical theory of aging' [1037; 1054]) ROSs are of course strongly implicated as partial contributors to the aging process (e.g. [6; 16; 872; 1037; 1055-1079]). Needless to say, not least because of the low steady-state net rate of generation of the various ROSs [1080] few studies have managed to be very specific mechanistically [1081], but it should be clear that all ROSs are not created equal and we need here to concentrate mainly on the 'nasty' parts of ROS metabolism, and in particular on the hydroxyl radical as generated via poorly liganded iron and on peroxynitrite, and to have the greatest effects we need to inhibit both their generation and their reactivity (see Systems Biology section, below). The iron content of cells also increases as cells age normally [1082]. As many diseases increase with age, probably via mechanisms highlighted herein, treating aging can thereby treat disease [120].

## Frailty

One issue of aging is not that just it happens but that it can manifest in a series of essentially undesirable physiological changes, referred to as frailty [1083], in which ROSs have also been strongly implicated. Indeed, there are many parts of physiology and metabolism that lose functionality during aging (e.g. the cardiovascular system [1084] and of course cognitive function [1085]), and iron metabolism is known to change considerably as humans age [872; 1086], with anaemia a typical accompaniment of aging [1087]. The question then arises as to how much of this deranged iron metabolism is causal in accelerated aging, and this is not easy to state at this time. However, lowering iron does increase the lifespan of *Drosophila* [1088] and yeast [989]. At all events, the purpose of this rather brief section is to point out to researchers in aging, frailty and gerontology generally the relevance of inadequately controlled iron metabolism <u>as a major part of</u> ROS-induced injuries that may accelerate the aging process.

## *Longevity*

Although aging and longevity are not of course the same thing, studies of aging are often performed with the intention of improving our understanding of longevity [1089], and certainly longevity is linked to age-related disease [1090]. However, the longevity of many organisms can be varied by the manipulation of any number of diseases or processes. This said, caloric or dietary restriction is a well-known contributor to longevity (indeed the only reliable one in pretty well all species [1091-1095] – although possibly not *H. sapiens* [1053]), and appears to act at the root of the processes involved [1096]. Caloric restriction appears to be associated with a considerably lowered rate of production of ROSs and accrual of ROS-induced damage [1080; 1097-1100], and this is to be expected on general grounds in that a restriction of substrate supply will make the redox poise of mitochondria [1101] more oxiding and thereby minimise the amount of 1- and 2-electron reductions of $O_2$ to form peroxide and superoxide. It is therefore also of great interest that caloric restriction also benefits iron status [1102] and that it is this improved iron status that in part promotes longevity [1103].

Antioxidants can also influence lifespan. Thus (rather high doses of) melatonin extended the lifespan (and stress resistance) of *Drosophila* [1064; 1104-1106] while that of *Caenorhabitis elegans* could be extended by mimics [1107] of SOD and catalase [1060], and by a variety of antioxidant and other pharmacological agents [1108].

As an example, let us consider *C. elegans*. Mutants with a decreased activity of the insulin-like growth factor signalling pathway (e.g. *daf2* mutants that have a greater amount of the DAF16 FOXO-like transcription factor) can live for nearly twice as long as wild types [1109; 1110] and produce more catalase, superoxide dismutase (sod-3) [1111] and glutathione-S-transferase.

Overall, it is the potent combination of oxidative stress, already leading to damaging peroxides and radicals, and its catalysis and further reactions caused by inappropriately chelated iron, that causes a 'double whammy'. Indeed, iron, copper and $H_2O_2$ have been referred to as the 'toxic triad' [923]. While there is little that we can do about the production of superoxide and peroxide, we can (by pharmacological or dietary means) try and improve the speciation of iron ions.

## Rheumatoid arthritis

One disease whose aetiology is well known to be bound up with ROSs is rheumatoid arthritis (RA) [10; 1112-1117]. What is known of the role of iron metabolism? Generally an overall low iron status – anaemia – is a characteristic of rheumatoid arthritis [1118-1122], whereas by contrast iron is elevated in the synovial fluid of arthritic joints [1123-1126]. This suggests a significant derangement of iron metabolism in RA as well, and a mechanism [1127-1130] in which elevated superoxide liberates free iron from ferritin in synovial fluid (and elsewhere [1131]), thereby catalysing further the damaging production of hydroxyl radicals. This autocatalytic process (Fig 6) is, even in principle, especially destructive (and may account for species differences in sensitivity to iron loading [1132]). Note that erythrocytes when oxidized can also release free iron [440]. Natural antioxidants such as vitamin E are also lowered [1133]. There is some evidence that appropriate iron chelators can ameliorate the symptoms of RA [1134], though membrane-impermeant chelators such as desferroxamine cannot [1135].

One interesting feature of RA is that in 75% of women it is strongly ameliorated during pregnancy [1136-1138], although the multifactorial nature of this observation has made a mechanistic interpretation difficult.

An interesting related feature is the 'restless legs syndrome' [1139-1141], that is often associated with iron deficiency and pregnancy. Serum transferrin receptor seems to be a rather sensitive measure of this iron deficiency [214; 1142; 1143]. There are relationships with other syndromes that we discuss here, such as cardiovascular disease [1144], but in the present review it is of especial interest that a dysregulation of iron metabolism appears to play a significant role [1145]

## Lupus (Systemic Lupus Erythematosus)

Lupus, or Systemic Lupus Erythematosus (SLE) [1146; 1147] describes a syndrome, somewhat related to arthritis and rheumatism, of broadly auto-immunological or inflammatory origin, and with a large variety of manifestations (e.g. [1148-1156]), but characterised in particular by fatigue [1157], often as a result of anaemia. This of course points to a certain level of iron dysregulation (of any number of causes), and there is certain some evidence for this [1145]. Thus, while anaemia is a feature of the disease, serum ferritin may be raised in SLE [1158], and some of the usual lipid markers of oxidative stress (that can be a result of hydroxyl radical production catalysed by poorly liganded iron) are also present [1159].

There is also a very interesting linkage betweem SLE and vitamin D metabolism [1160], something that has also come up in relation to the statins and atherosclerosis ([725; 1161], but cf. [1162]), and indeed it there is evidence that statins may be of benefit in the treatment and even reversal of lupus [1163; 1164]. Indeed, from a much more general point of view, there is precedent for these kinds of linkages being used to uncover unknown mediators in disease states that may be worth pursuing (e.g. [1165-1169]).

### Asthma

Asthma is a well-known inflammatory disease, and has been linked with ROS generation [1170] catalysed by iron [217; 1171-1173].

### Inflammatory bowel diseases (IBD)

By definition, IBD such as Crohn's disease and ulcerative colitis are inflammatory diseases, and while the inflammation and ROS production are well established here, their origins are somewhat uncertain [1174-1177]. They are frequently accompanied by anaemia, implying a derangement in iron absorption and/or metabolism [1178], and very probably absorption [1179; 1180]. The anaemia may be monitored by ferritin and transferrin receptor levels, and its correction is possible by iron supplementation plus erythropoietin [1181-1187].

### Age-related macular degeneration.

Age-related macular degeneration (AMD) [1188] is now the leading cause of blindness and visual disability in the elderly in developed countries [1189-1192]. Many components of atherosclerotic plaques have also been demonstrated in drusen [1193], a characteristic of AMD and, as here, it is reasonable to propose a common mechanism of pathogenesis between AMD and atherosclerosis. Retinal iron levels increase with age [1194], iron is significantly implicated in AMD [1195-1201], and iron chelation may help to reverse the process [1202]. The source of the iron appears to be excess angiogenesis and leakage from blood vessels catalysed by VEGF, and a PEGylated aptamer [1203-1205] against VEGF (pegaptanib) or a monoclonal antibody (ranibizumab) have shown significant promise in the treatment of macular degeneration [1206-1211]. Plausibly a combination therapy with one of these plus a suitable iron chelator might be even more effective.

### Psoriasis

Psoriasis is an anflammatory disease in which the production of free radicals and ROSs are strongly implicated [1212-1214]. Here too there is clear evidence for the involvement of a deranged iron metabolism [1214-1216]. Early attempts at therapy with a series of unusual iron chelators (that unfortunately had side effects) [1217] do not seem to have been followed up.

### Gout

Gout is another important inflammatory disease, characterised by the accumulation of uric acid [1218]. There is considerable evidence that this too is a disease of iron overload, and that uric acid accumulation – as both an antioxidant and an iron chelator [1219] – is in response to the iron overload [1220; 1221] and with highly beneficial remission of gouty symptoms occurring on depletion of iron by phlebotomy [1222].

## Alcoholic and other forms of liver disease

It is known that with chronic excess, either iron or alcohol alone may individually injure the liver and other organs, and that in combination, each exaggerates the adverse effects of the other. Specifically, in alcoholic liver disease, both iron and alcohol contribute to the production of hepatic fibrosis [1223-1228]. Iron overload is well known to lead to hepatotoxicity [1229; 1230] and liver cancer [1231; 1232], and lowering or chelating it is protective [1233; 1234]. Hepcidin may be involved here [1235].

## Chronic obstructive pulmonary disorder (COPD) and related lung diseases

Chronic obstructive pulmonary disease (COPD) is a progressive and chronic disease which is characterised by an apparently inexorable decline in respiratory function, exercise capacity, and health status. It is also characterised by periods in which the symptoms are considerably exacerbated [1236-1239]. Such an acute exacerbation of COPD (AECOPD) is defined [1240] as "a sustained worsening of the patient's condition from the stable state and beyond normal day to day variations, that is acute in onset and necessitates a change in regular medication in a patient with underlying COPD''. Smoking is a major source of free radicals (and indeed metals), and is a major cause of COPD. Consequently, there is considerable evidence for the evidence for the involvement of inflammation and ROSs in both the 'stable' and 'exacerbated' stages [1241-1248].

Needless to say, there is also considerable evidence for the significance of exposure to iron [1249] (as well as exposure to other toxic metals [1250]) in the development of COPD and other lung diseases [1251]. Haem oxygenase also appears to be a significant culprit [221], and lung (lavage) iron is increased [1252] while transferrin levels can be considerably lower [1253].

Other lung diseases in which ROS and iron have been implicated include Adult Respiratory Distress Syndrome [1254-1257] and cystic fibrosis [1258].

## Smoking

Tobacco smoke contains many unpleasant and carcinogenic compounds, and that tobacco smoking is a leading cause of carcinoma of the lung and indeed of other organs has become well known since the pioneering epidemiological studies of Doll, Peto and colleagues (e.g. [1259; 1260 ; 1261; 1262]). Our purpose here is to point out that many of the particles associated with smoking and ingested from other sources are heavily laden with micro-particulate iron, which, as a major catalyst of hydroxyl radical production, undoubtedly is a substantial contributor as well (see e.g. [1251; 1263-1267]).

## Cancer and oncology

In addition to the issues of smoking, the development of cancer can certainly contain an inflammatory component [1268-1278], and indeed the long-term use of prophylactic anti-inflammatory aspirin lowers colon cancer incidence by 40% (age-adjusted relative risk = 0.6) [1279]. Given that cells require iron, restricting its supply can also limit the growth of cells, including tumour cells [1280-1288]. Conversely the iron carrier NGAL is overexpressed in tumours [330; 331; 1289]. Further, the roles of iron, not least in the mutagenic effects of metal-catalysed Fenton chemistry, are also of significance in promoting oncogenesis [1290-1305]. Iron chelators [1287] are thus a doubly attractive component of anti-cancer therapeutics. The mutagenic, carcinogenic and disease-causing actions of asbestos and related fibres may also be due in significant measure to the ability of the Fe(n) that they contain to catalyse hydroxyl radical production [1306-1318], while there seem to be complex relations between the likelihood of apoptosis and the differential concentrations of superoxide and $H_2O_2$ [1319].

## Malaria

Just as do tumour cells, the malarial parasite *Plasmodium falciparum* requires considerable iron for growth, and there is evidence that lowering the amount of available iron provides a promising route to antimalarials (e.g. [1320-1336], but cf. [1337]). Note in this context that iron-catalysed radical formation is also significantly involved in the antimalarial (i.e. cytotoxic) mode of action of artemisinin [1336; 1338-1346], and this reaction is in fact inhibited by iron chelators [1347] such that a combined artemisinin-chelator therapy would be contraindicated.

## Antimicrobials

Lower down the evolutionary scale, and as presaged earlier in the section on bacterial siderophores, microbes require iron for growth, its presence may be limiting even at the scale of global $CO_2$ fixation [1348-1351], its excess can in some circumstances correlate with infectivity or virulence (see above and e.g. [136; 149; 1335; 1352-1371]), and its chelation in a form not available to bacteria offers a route to at least a bacteriostatic kind of antibiotic or to novel therapies based on the lowering of iron available to microbes by using hepcidin [1372]. Iron chelators are also inhibitory to trypanosomes [1373; 1374].

## Sepsis leading to organ failure and death: severe inflammatory response syndrome

It is well known that one consequence of bacterial infection (sepsis) can be septic shock, that this can be mimicked by the Gram-negative bacterial outer membrane component LPS (lipopolysaccharide), and that in the worst cases this leads via multiple organ failure to death. However, whatever LPS does it is quite independent of the present of viable (i.e. growing or culturable – see [1375; 1376]) bacteria as the same phenomena leading to multiple organ failure (MOF) are seen without infection [1377; 1378]. Consequently, the recognition of a series of symptoms contingent on this initial inflammatory response has led to the development of the idea of a Systemic Inflammatory Response Syndrome (SIRS) [1379-1389] that leads to the MOF, both via apoptosis [1390] and necrosis [1391; 1392]. There is by now little doubt that these phenomena too are associated with the hyperproduction of ROSs [1384; 1389; 1393-1408]. Circulating free iron is raised in sepsis and related conditions [1267; 1409-1411]. Direct assays of oxidant induced cell death indicate that most 'free' iron is concentrated in lysosomes [1412-1415], that its decompartmentation is substantially involved [1408], and that its chelation can thus prevent cell death [1416-1419].

Many circulating inflammatory factors have been identified as important in the development of septic shock, including cytokines such as Tissue Necrosis Factor (TNF) [1420], and cellular responses via the Toll-Like Receptor are clearly involved in this process [1421; 1422]. However, we would argue that since antibodies against TNF do not inhibit the sequelae of septic shock such as multiple organ failure, the truly damaging agents are caused elsewhere and are likely to involve the iron-mediated production of damaging hydroxyl radicals (see also [1401]).

In this regard, it is especially interesting that the antioxidant melatonin is particularly effective in preventing septic shock [1423-1425], and a variety of suitable antioxidants have shown potential here [1411; 1426-1429], notably in combination with iron chelators [1430; 1431] (and see also [1432; 1433]). As with quite a number of the indications given above, a further link with Fe metabolism is seen in the protective effects of erythropoietin [1434; 1435].

# Pro- and anti-oxidants and their contributions to cellular physiology

A very great many cellular metabolites are redox active within the range of redox potentials realistically accessible to biology (including some molecules such as proline [1436] that are not commonly considered to be so), and it is not our purpose here to list them extensively. Not only their redox potential and status but even their absolute amounts can have profound effects on metabolism [1437]). Our chief point here is that it is the intersection of iron metabolism and oxygen reduction that needs to be the focus, with the 'iron'-catalysed production of hydroxyl radical being the nexus, and with the absolute redox potential of a redox couple *per se* being less significant in absolute terms, and the redox potential that a particular redox couple 'feels' being dependent in a complex manner on a variety of thermodynamic and kinetic factors [1101]. Thus, although ascorbate is 'reducing' and an 'antioxidant', its reaction with $O_2$, especially when catalysed by Fe(II), produces superoxide and thence $OH^\bullet$ radicals that may be pro-oxidant. It is this kind of stepwise multi-electron-transfer phenomenon that explains the otherwise possibly puzzling observation of the oxidant-induced reduction of respiratory chain components (see e.g. [1438; 1439]). Consequently, it is extremely unwise to make pronouncements on the role of 'ROSs' without being quite explicit about which ones are meant.

Thus anything – even an antioxidant – that e.g. by reaction with $O_2$ produces superoxide, peroxide and hydroxyl radicals will turn out to be a pro-oxidant if the flux to superoxide and in particular to hydroxyl radicals is stimulated. Thermodynamically, the 1-electron reduction by ascorbate of dioxygen is disfavoured, with the 2-electron reduction to peroxide being the thermodynamically preferred route. However, such reactions are heavily restricted kinetically in the absence of any catalysts [95]. It is an unfortunate fact that the oxygen-mediated "autoxidation" of ascorbate does in fact occur at considerable rates when it is accelerated by the presence of iron or other transition metal ions [95; 1440-1443]. In a similar vein, 'free' or inadequately liganded Fe(II) catalyses the production of hydroxyl radicals from oxygen plus a variety of natural biomolecules, including adrenaline (epinephrine) [1444], haemin [1445], and even peptides such as the amyloid-β involved in the development of Alzheimer's disease [900; 918]. Dietary antioxidants (see below) can therefore act in complex and synergistic ways depending on iron status [1446]. In this regard, the idea of using elemental iron plus ascorbate in food supplements [1447] does not seem a good one.

It should be noted that there are also occasions, e.g. in the decomposition of refractory polymers such as lignin, where such radical production is involved beneficially [1448].

Finally, a variety of molecules can trap hydroxyl radicals, including hippurate [1449], melatonin [1104; 1423; 1424; 1450-1467] and salicylate [1468].

# Antioxidants as therapeutic agents? Should we be including iron chelators in such clinical trials?

Given the wide recognition of the importance of ROSs in a variety of diseases as described above, many investigators have considered the use of known antioxidants such as vitamins C (ascorbate) and E (α-tocopherol) in preventative therapy. Although there have been some successes (e.g. [1469]), the results have generally been decidedly mixed, with little clinical benefit (or even actual disbenefit) following from their administration [600; 1426; 1470-1475], e.g. for ALS [1476], atherosclerosis [76; 669], cardiovascular disease [1477-1481], neuroprotection [1482], macular degeneration [1483], pre-eclampsia [1484-1488], critical care medicine [1489], aging [1039; 1042; 1490; 1491], lung disease [1492], elective surgery involving ischaemia-reperfusion [1493], all-cause mortality [1473; 1494], etc. One interpretation for these disappointing results that is consistent with the general theme of this review involves the recognition that a variety of <u>anti</u>oxidants can act as <u>pro</u>-oxidants and thus actually <u>promote</u> the production of OH<sup>•</sup> radicals in the presence of inappropriately or inadequately liganded Fe(II) [95; 444; 1440; 1442; 1495; 1496]. (One might also comment that the intracellular location of the antioxidants may be an issue, and the view that targeting them to mitochondria may well have considerable merit [1497; 1498].) Thus any predictions about the utility or otherwise of antioxidants need to take into account the amount of 'free' iron present. **In particular, we would suggest that future trials of this type might beneficially include appropriate iron chelators, whether alone <u>or</u> with antioxidants.**

# Liganding and reactivity of Fe(n)

Given the damage that iron-mediated OH<sup>•</sup> radical can create, the question arises as to whether appropriate chelators can inhibit this by inhibiting the production of OH<sup>•</sup>, and while the answer is 'yes' the interpretation of the relevant experiments is not always as clear cut as one would wish [16]. This is because the OH<sup>•</sup> radical is so reactive that its production is normally assessed by addition of the putative chelator and observation of its effect on the rate of reaction of a target molecule such as salicylate with the OH<sup>•</sup> generated. The ability of a chelator to inhibit such a reaction can then occur not only via a reduction in the rate of OH<sup>•</sup> production but by trapping the OH<sup>•</sup> itself, as well as by other mechanisms [1499]. This said, there is little doubt that iron chelators can be highly protective, and it is many ways very surprising that their use is not more widespread.

We begin by noting that the reactivity of iron does vary greatly depending upon its liganding environment [41]. Cheng et al. state [42] "Oxygen ligands prefer Fe(III); thus, the reduction potential of the iron is decreased. Conversely, nitrogen and sulfur ligands stabilize Fe(II); thus, the reduction potential of the iron is increased. Therefore, chelators with oxygen ligands, such as citrate, promote the oxidation of Fe(II) to Fe(III), while chelators that contain nitrogen ligands, such as phenanthroline, inhibit the oxidation of Fe(II). Many chelators, such as EDTA and Desferal (DFO), will bind both Fe(II) and Fe- (III); however, the stability constants are much greater for the Fe(III)-chelator complexes. Therefore, these chelators will bind Fe(II) and subsequently promote the oxidation of the Fe(II) to Fe(III) with the concomitant reduction of molecular oxygen to partially reduced oxygen species. Since the maximal coordination number of iron is six, the hexadentate chelators can provide more consistently inert complexes due to their ability to completely saturate the coordination sphere of the iron atom and, consequently, deactivate the "free iron" completely. For example, DFO is a very effective antioxidant in clinical application because of its potential to markedly decrease the redox activity of iron [102]." However, it is not each to make hexadentate ligands orally active [1500].

Iron typically can coordinate 6 ligands in an octahedral arrangement. Preferential chelation of the Fe(II) or the Fe(III) form necessarily changes its redox potential as a result of Le Chatelier's principle, and from Marcus theory [1501-1504] the rate of outer-sphere electron transfer reactions is typically related to differences in the free energy change, i.e. the differences in redox potentials of the interacting partners. In addition, it widely recognised that [102] "The tight binding of low molecular {weight} chelators via coordinating ligands such as O, N, S to iron blocks the iron's ability to catalyze redox reactions. Since the maximal coordination number of iron is six, it is often argued that the hexadentate chelators can provide more consistently inert complexes due to their ability to completely saturate the coordination sphere of the Fe atom. Consequently, a chelator molecule that binds to all six sites of the Fe ion completely deactivates the "free iron". Such chelators are termed "hexidentate" {sic}, of which desferrioxamine is an example. There are many Fe chelators that inhibit the reactions of Fe, oxygen, and their metabolites. For example, desferrioxamine … (DFO) markedly decreases the redox activity of Fe(III) and is a very effective antioxidant through its ability to bind Fe."

By contrast, bidentate or tridentate chelators that bind to only 2 or 3 of the available iron chelation sites, especially when they bind to both Fe(II) and Fe(III), can in fact catalyse redox cycling and thereby promote free radical generation [1286; 1500; 1505; 1506]. Thus, the most potent iron chelators will normally be hexadentate  (but may consequently strip iron from iron-containing enzyme and thereby have deleterious side effects). Thus bi- or tri-dentate ligands should be at saturating concentrations for maximum effect.

Generally, the harder ligands that favour Fe(III) involve O whereas softer ligands that bind Fe(II) involve N and S. The type of ligand also influences the absorption spectrum of the ferric form of the chelator, such that conclusions can be drawn about the types of group involved in the complex. These charge transfer bands that appear on ligand binding are at around 340 nm for carboxylates, around 425 nm for trishydroxamates, 470 nm for bis-hydroxamates, 515 nm for monohydroxamates, around 480 nm for tris-catecholates, 560 nm for bis-catecholates and 680 nm for mono-catecholates [129]. In addition, for tris-bidentate complexes the complex can, on an octahedral arrangement, have two different configurations, a left-handed propeller, termed the Λ-configuration, and a righthanded propeller, the Δ-configuration [129].

## *Iron chelators – those approved and used clinically*

A number of reviews (e.g. [1286; 1287; 1507-1513]) cover aspects of iron chelators that have had or may have utility clinically.

Whitnall and Richardson [952] list a number of useful features of an experimentally (and clinically) useful iron chelator. Thus, "A compound suitable for the treatment of neurodegenerative disease should possess a number of qualities, namely (1) strong affinity for FeIII, (2) low molecular weight, (3) lipophilicity high enough to accommodate permeation of cell membranes and the BBB, (4) oral activity, and (5) minimal toxicity [952]. Also, partly because there are few trivalent ions other than Fe(III) that the body actually needs, the major synthetic focus has been on the design of FeIII-selective chelators which feature "hard" oxygen donor atoms. Additionally, under aerobic conditions, high-affinity FeIII chelators will tend to chelate FeII to facilitate autoxidation, such that high-affinity FeIII-selective compounds will beneficially bind both FeIII and FeII under most physiological conditions" [952].

Desferrioxamines are nonpeptide hydroxamate siderophores composed of alternating dicarboxylic acid and diamine units. linked by amide bonds. They are produced by many *Streptomyces* species [1514]. Desferrioxamine B is a linear (acyclic) substance produced (industrially) by the actinobacterium *Streptomyces pilosus* [1515], and is widely used as an iron chelator for the prevention and treatment of the effects of iron overload. It is commercially available as desferal (desferrioxamine methane sulphonate), also known as deferoxamine in the USA. It has been very effective in the treatment of a number of diseases, leading to the view that such molecules should have considerable therapeutic potential. A significant disadvantage of DFO is that it does not seem to cross the intestine intact (despite the rather catholic substrate specificity of intestinal peptide transporters [1516-1519]) and must therefore be given intravenously or subcutaneously. By contrast, another chelator known as Deferriprone or L1 does appear to cross cell membranes, but it is only bidentate.

Those with approval for clinical use are few in number and we deal with them first. Table 1 compares them with the 'ideal' properties of a clinically useful iron chelator, while Fig 7 gives the structure of the three most common, viz. desferal (deferoxamine), ferriprox (L1 or deferiprone) and exjade (ICL670 or deferasirox). (Dexrazoxane, a hexadentate chelator, is also marketed [1520]. )

Desferal (deferoxamine) is the most used chelator for historical reasons. It is hexadentate but is not orally bioavailable. Ferriprox (deferiprone) is a bidentate ligand (1,2-dimethyl, 3-hydroxypyridin-4-one). It is orally bioavailable although comparatively high doses are required, and postdates . "Deferiprone has high affinity for iron and interacts with almost all the iron pools at the molecular, cellular, tissue and organ levels. Doses of 50–120 mg/kg/day appear to be effective in bringing patients to negative iron balance" [1521]. It can have somewhat better properties than desferal [1522]. Finally, Exjade (ICL670) (deferasirox) [1523-1533] is the most recent chelator approved for clinical use, and is tridentate. There is a large bibliography at **http://www.exjade.com/utils/bibliography.jsp?disclaimer=Y.** The recommended initial daily dose of EXJADE is 20 mg/kg body weight.

It is clear from Table 1 that in the time evolution from deferoxamine through deferiprone to deferasirox there has been a noticeable improvement in the general properties of iron chelators, although there are few published data on the quantitative structure-activity relationships of candidate molecules that might allow one to design future ones rationally. What is certainly clear is that there is a trade-off in properties, and that appropriate chelators will keep iron levels intermediate, i.e. not too low and not too high (a 'Goldilocks' strategy, if you will), and that hexadentate molecules may correspondingly be too tightly binding and strip iron from important molecules that need it. What is particularly important, as well as a good plasma half-life, is the ability to cross cell membranes, as this is necessary both for oral administration and for ensuring that the chelator in question actually accesses the intracellular 'free' iron pools of interest. Which carriers are used for this in humans in vivo is presently uncertain [1534].

### *Drugs that have been approved for clinical use for other purposes, but that also happen to be iron chelators*

The high investment of time, money and intellectual activity necessary to get a drug approved clinically has led to a number of strategies to exploit those that already have been approved and are thus considered 'safe'. One such strategy is the combination therapy of approved drugs that can yet serve for novel indications (e.g. [1535-1539]) . Another strategy is to look for antioxidant or iron-binding chemical motifs in drugs that have already been approved for other purposes [1540] (or to measure such properties directly). This has led to

Clioquinol (CQ) [952; 1509; 1541; 1542] (Fig 7) is one existing (anti-parasitic) drug that has been proposed for use as an iron chelator, as it contains the known iron-chelatiing 8-hydroxyquinoline moiety. It has indeed enjoyed some success in this role. However, clioquinol toxicity has been reported if it is used over an extended period [1543] and this may be due to the formation of a Zn-clioquinol chelate [1544].

A particular attraction of such existing drugs is that they are likely to have favourable pharmacokinetics and pharmacodynamics, and in particular are likely to be cell-permeable. Note that despite a widespread assumption that lipophilicity or log P is sufficient to account for drug distribution this is not in fact the case, as there are hundreds of natural transporters that drugs can use (e.g. [1534; 1545]). For instance, the iron-chelating 8-hydroxy quinoline motif contained in molecules such as clioquinol is also present in the tryptophan catabolite xanthurenic acid (Fig 7), and it is likely that transmembrane transport of the synthetic drug molecule occurs via natural carriers whose 'normal' role is to transport endogenous but structurally related molecules.

## Iron chelators that have been studied but not yet approved

Given the importance of the field, many academic investigators have sought to develop their own iron chelators that might exhibit the desirable properties listed above. One class of molecule includes isonicotinylhydrazones. Thus, pyridoxal isonicotinyl hydrazone (PIH) [1507; 1546-1551] is a promising molecule (also proposed in anti-cancer therapy), although it is hydrolysed both *in vivo* and *in vitro* [1552]. Other analogues include salicylaldehyde (SIH) [1553] and 2-hydroxy-1-napthylaldehyde (NIH) isonicotinyl hydrazones. PIH was disclosed before being patented, and is thus seen as having no pharmaceutical (company) interest. Various other derivatives are therefore being considered [926], including pyrazinylketone isonicotinoyl hydrazones [1554].

A variety of 8-hydroxyquinolines (8HQs) [1555] have been considered, although as with other bidentate and tridentate ligands that cannot necessarily effect complete liganding of iron there is always a danger that inadequate concentrations might be pro-oxidant (e.g. [1556; 1557]). One in particular, VK-28, combines various pertinent moieties and has shown some promise in the treatment of neurological disorders [1558-1561]. This strategy of combining drug elements that can hit multiple targets ('polypharmacology' [1562-1564]) has much to commend it, including on theoretical grounds, and we discuss these in the section on systems biology below. Another 8HQ that has elicited interest is O-trensox [1280; 1565-1572].

Other ligands or motifs that might be considered include di-2-pyridylketone-4,4,-dimethyl-3-thiosemicarbazone (Dp44mT), that has been shown to be effective against tumours [1573], 2,2'-dipyridyl, 1,10-phenanthroline [1574; 1575], 2-benzoylpyridine thiosemicarbazones [1576] and thiohydrazones [1577]. HBED (Fig 7) (N,N′-bis-(2-hydroxybenzyl)ethylenediamine-N,N′-diacetic acid) forms a 1:1 complex with Fe(III) but is probably only tetradentate [1578]. It seems not to be very orally active [1579] but may be more effective than is DFO [1580-1582]. Poly-hydroxylated 1,4-naphthoquinones occur as sea urchin pigments and have shown protective effects [1583].

Continuing the theme of polypharmacology, R-(α)-lipoic acid [1584-1587] is also an antioxidant. Finally, one interesting area is that of prochelators (e.g. [1588]) in which the oxidant itself triggers the formation of a chelator able to inhibit the Fenton reaction.

## Utility of iron chelators in disease amelioration

Therapeutic uses of iron chelators have been widely and usefully reviewed (e.g. [1286; 1331; 1511; 1589-1596]. Many problems remain, such as bioavailability, mis-dosing [1597] (too little iron as well as too much of it can be bad), toxicity, selectivity and so on, and their design is consequently highly non-trivial [1500; 1505]. Nevertheless, iron chelators have demonstrated therapeutic benefits in Alzheimer's [1509; 1541; 1598-1600], Parkinson's [928; 1560; 1601], cold-induced brain injury [1602; 1603], coronary disease [621; 694], renal diseases [1604], various kinds of infection [1592] and of course in iron overload diseases [1591; 1596].

As mentioned above, one interesting strategy is to devise chelators that are only activated by oxidative stress [1589; 1605-1608]. Another is to seek to combine different kinds of functionality in the same molecule. To this end, Youdim and colleagues in particular have developed a series of multifunctional 8-hydroxyquinoline [1570] derivatives that are effective bidentate iron chelators and that seem to show considerable promise in the treatment of a variety of neurodegenerative diseases [928; 1558-1560; 1609-1613] (see also US Patent 20060234927). In this case the antioxidative mechanism is clearly via chelation since such (8-hydroxyquinoline) molecules are poor scavengers of radicals directly [1614], a fact that also makes them useful scientific tools. As bidentate ligands they cross both cell membranes and the BBB fairly easily (though lipophilicity *per se* seems not to be important for the biological activity of 8-hydroxyquinoline chelators [1567; 1569]). Importantly, the comparatively weak bidentate binders seem not to have major long-term effects if used carefully [1325; 1591; 1615; 1616]

## Interaction of xenobiotics with iron metabolism

As Cherny and colleagues point out [1541], there are many US Pharmacopaeia-registered drugs that, while not being termed chelators, do in fact have both chelating properties and favourable toxicity profiles. Thus we need to recognise potentially both positive and negative interactions between drugs in general and iron metabolism. Any drug that can bind iron can also catalyse the formation of free radicals. Thus, gentamicin can form a gentamicin-iron complex that can lead to toxic symptoms such as hearing loss; this is reversed by iron chelators [1617; 1618]. Existing drugs other than iron chelators may also have effects on iron metabolism [1619], and iron can catalyse their oxidation [1620]. It is not, of course, news that drugs have multiple effects. In this context, we reiterate that some statins, for instance, have chelating properties [786].

Other toxicants might also mediate their damaging effects through iron-catalysed radical formation [1621; 1622]. This in addition to the well-known iron-catalysed, radical-mediated mechanism of toxicity of the viologens such as diquat and paraquat [1623-1628] (whose herbicidal activity is in fact inhibited by iron chelators [1629]) and of adriamycin [1630; 1631]. As mentioned above, the carcinogenic action of asbestos may also be due to the ability of the Fe(n) that it contains to catalyse hydroxyl radical production [1312; 1318], while carcinogenic mycotoxins such as aflatoxin may interact synergistically with iron [1632].

### *Dietary sources of iron chelators*

There is also a considerable and positive role for nutrients in terms of their chelation of iron. Indeed, polyphenolic compounds, many of which have known health benefits [1633-1642], are widely used as food antioxidants [1643; 1644]. There is of course considerable epidemiological evidence for the benefits of consuming fruit and vegetables that are likely to contain such antioxidants (e.g. [1645-1648]), and – although possibly a minimum – this has been popularised as the 'five a day' message (e.g. http://www.fruitsandveggiesmatter.gov/ and http://www.5aday.nhs.uk/). Even though elements of the 'Mediterranean' diet that are considered to be beneficial are usually assumed to be so on the basis of their antioxidant capabilities (but cf. [1649]), many of the polyphenolic compounds (e.g. flavones, isoflavones, stilbenes, flavanones, catechins (flavan-3-ols), chalcones, tannins and anthocyanidins) [1650-1657] so implicated may also act to chelate iron as well [963; 1658-1672]. This is reasonable given that many of these polyphenols and flavonoid compounds [1650; 1673-1682] have groups such as the catechol moiety that are part of the known iron-binding elements of microbial siderophores. Examples include flavones such as quercetin [807; 1642; 1658; 1683-1693], rutin [1658; 1686; 1687; 1694; 1695], baicalin [1689; 1696], curcumin [1642; 1697-1701], kolaviron [1702], flavonol [1703], floranol [1704], xanthones such as mangiferin [1705-1708], morin [1705], catechins [963; 1636; 1667; 1683; 1709; 1710] and theaflavins [1711], as well as procyanidins [1664; 1712] and melatonin [1464; 1713-1716]. However, the celebrated (*trans*-)-resveratrol molecule [1717-1731] may act mainly via other pathways.

A considerable number of studies with non-purified dietary constituents containing the above polyphenolic components have also shown promise in inhibiting diseases in which oxidative stress is implicated [1654; 1732-1734]. For instance in stroke and related neuronal aging and stress conditions, preventative activity can be found in blueberries [1735-1741] (and see [1742]), *Ginkgo biloba* extract (EGb 761) [1738; 1743; 1744], grapes [1745], green tea [1636; 1746-1749], *Mangifera indica* extract [1708], strawberries [1735], spinach [1735] and *Crataegus* [815], while combinations of some these components ('protandim') have been claimed to reduce ROS levels by stimulating the production of catalase and SOD [1750]. As with pharmaceutical drugs [1534; 1751-1753], there are significant problems with bioavailability [1754; 1755], although the necessary measurements are starting to come forward [1633; 1638; 1754-1760]. There is now increasing evidence for the mechanisms with which these dietary components and related natural products and derivatives (often with anti-inflammatory, anti-mutagenic or anti-carcinogenic properties) interact with well recognised cellular signalling pathways (e.g. [1268] [1761] [1762] [1763] [1764] [1765] [1766] [1767] [1768] [1769] [1770] [1771] [1772] [1724] [1773] [1774] [1775] [1776] [1777] [1778] [1779] [1780] [1725] [1273] [1781] [1782] [1783] [1784] [1785] [1786] [1787] [1788] [1276] [1789] [1790] [1791] [1792] [1793] [1794] [1795] [1796] [1797] [1798] [1799] [1800] [1801] [1802] [1803] [1804] [1805] [1806] [1807] [1808] [1809] [1741] [1810] [1729] [1811] [1812] [1813] [1814] [1815] [1816] [1817] [1818]).

# Role of iron-generated ROSs in cellular signalling and oxidative stress

Thus, although this is not the focus of the present more physiologically based review, we recognise that many of relationships between ROSs and oxidative stress and overt progressive diseases may be mediated via the inflammatory signalling pathways involved in 'innate immunity' [793; 1819-1821]. The NF-κB system is intimately involved in this [503; 580; 625; 1271; 1272; 1299; 1822-1847].

In the NF-κB system (e.g. [1848-1852]) (Fig 8), NF-κB is normally held inactive in the cytoplasm by binding to an inhibitor IκB (often IκBα). Pro-inflammatory cytokines such as TNF-α, LPS [1853-1858] and IL-1 [1859] act by binding to receptors at the cell surface and initiating signalling pathways that lead to the activation of a particular kinase, IκB kinase or IKK. This kinase phosphorylates the IκB causing it to be released (and ubiquitinated and degraded by the proteasome), allowing the NF-κB to be translocated to the nucleus where it can activate as many as 300 genes. Simple models of the NF-κB system show the main control elements [1860; 1861] and their synergistic interaction [1862]. The NF-κB system is implicated in apoptosis [1863; 1864], aging [1072], and in diseases such as cancer [1269; 1299; 1637; 1865-1868], arthritis [1868-1871] and a variety of other diseases [1872]. Antioxidants such as vitamin E [467; 1845] and melatonin [1873-1877] are at least partially protective. Oxidative stress seems to act upstream of IKK [1842], on IκBα directly [1878] and in the p38 MAP kinase pathway [1821; 1842; 1879], and there is also evidence that at least some of the statins act on the PI3K-akt and NF-κB pathways too [712; 776; 1880-1887].

The induction of NF-κB by ROSs appears to involve a coupling via the glutathione system [1835; 1863; 1864; 1888-1905] (and see also [1906; 1907]).

A variety of studies have shown that iron is involved in these signalling processes [1668; 1824; 1843; 1908-1912], probably again acting upstream of the IKK [472; 1910; 1913; 1914].

Interestingly, there is interplay between the NF-κB pathway and the regulation of NGAL [324; 325; 1915], ferritin [1858] and hepcidin [1916], presumably acting as a negative feedback as the cell tries to control and ligand its free Fe(n) in the face of oxidative stress caused by the release of free iron [1917].

## The systems biology of ROS generation and activity

It is not news that most major changes in physiological states have multigenic or multifactorial origins (e.g. [1918; 1919]). This means, as an inevitable consequence, that we need to recognize that their observation requires a systems approach, and that most diseases are therefore in fact to be seen as systems or network diseases [541; 1920-1928]. Changes in individual reaction steps (or even single manipulations) can changes the levels of scores or hundreds of transcripts [1929], proteins [1930] or metabolites [631]. In this regard, small molecule (metabolomics) measurements have especial advantages for capturing network organisation, including on theoretical grounds [1931-1938].

If we consider just one variable of present relevance, the quantity of hydroxyl radical, the amount that is able to react with proteins, lipids and DNA is clearly determined by a huge number of reactions, whether directly or otherwise – not only the concentrations of reagents and enzyme that directly catalyse its formation and destruction but by everything else that affects their concentration and activity, and so on. This is of course well established in biology in the formalism of metabolic control analysis (MCA) [1939-1942], and was recognized over 30 years ago in Morris' prescient review on anaerobes [1943]. Modern systems theories of aging (e.g. [1049; 1053; 1944]) (Fig 9) also recognize physiological progression as being determined in terms of a balance between 'good' and 'bad' factors.

MCA and related formalisms can be seen as theories of sensitivity analysis, which in many cases can be normalized such that an overall output function can be described <u>quantitatively</u> in terms of the relative contributions of each of its component steps (e.g. [1945-1949]). In MCA the normalized local sensitivities are known as control coefficients, and the sum of the concentration-control coefficients = 0, in other words in the sateday state the rate of production and consumption of a particular entity is in balance and <u>all</u> reactions can contribute to it to some degree. The concentration-control coefficient describes this degree quantitatively. It is now possible to produce appropriate quantitative representations of metabolic networks using quite sparse kinds of information (in fact just the stoichiometry and network structure [1950]), and thereby provide initial estimates for more sophisticated fitting algorithms (e.g. [1951-1954]. Indeed, the analysis of the properties and behaviour of networks is at the core of modern systems biology (e.g. [1921; 1955-1959]).

A corollary of such considerations is that to decrease the amount of damage caused by OH$^{\bullet}$ (or any other) radicals we need both to decrease their production and increase their removal to harmless substances [1960], and that on general grounds [1537-1539; 1961] such a strategy (for instance of combining a cell-permeable iron chelator with a cell-permeable antioxidant) might be expected to give a synergistic response. Even determining the means of cell permeability and tissue distribution turns out to be a systems biology problem in which we need to know the nature and activity of all the carriers that are involved [1534; 1545; 1962]. At all events, it is undoubtedly the case that the steady-state rate of production of a molecule such as the hydroxyl radical is controlled or affected by a considerable number of steps. These minimally includes the multiple reactions of the mitochondrial respiratory chain and the various oxidases producing superoxide and peroxide, the activities of catalase and SOD enzymes that together can remove them, protective reactions such as heat-shock proteins, and most pertinently to the present review a large number of reactions involved in the metabolism and safe liganding of iron that help determine the rate at which OH$^{\bullet}$ is produced.

It is also pertinent to enquire as to why we are now seeing so many of these progressive diseases, and as to what may be their causes. Undoubtedly the simple fact of improved longevity is one [123] as damage accumulates. However, anything that decreases the amount of unliganded iron, such as decreasing the total dietary iron intake e.g. from red meat, must be helpful [1033; 1963].

## *Anti-inflammatory cytokines; the example of erythropoietin*

We have above adduced considerable evidence that decreasing the amount of hydroxyl radical by any means is valuable, whether by removing initially generated ROSs such as superoxide and peroxide or by chelating poorly liganded iron in a way that stops these ROSs forming the hydroxyl radical. While pro-inflammatory cytokines can themselves increase ROS production and modulate the activities of signaling pathways such as NF-κB and p38, there are also anti-inflammatory cytokines. A particularly interesting example is that of erythropoietin.

Erythropoietin was originally recognized via its role in erythropoiesis [1964-1966] (hence its name, of course), but it has become evident that it has many other roles, and in particular it is observed phenomenologically that erythropoietin (and non-erthyropoetic derivatives) is protective in a number of inflammatory conditions that accompany many diseases such as those listed above [832; 834; 1967-1974]. These included cardiovascular disease [627; 628; 1975-1992], stroke and other related neurological diseases [817-819; 821; 822; 826; 829; 830; 1974; 1993-2020], diabetic neuropathy [2021], kidney injury [1992; 2022-2028], intestinal injury [2029] and shock (both septic and non-septic) [1434; 1435; 2030].

The question then arises as to how it is doing this mechanistically, and the proximate answer is that it (and other anti-inflammatory agents, e.g. [1637; 2031; 2032]) seem to act via many of the same signalling pathways as do inflammatory agents [835; 1969; 2033-2042]. There is evidence that it can help maintain superoxide dismutase activity [2030; 2043], invoke haem oxygenase [2044], and in particular – from the perspective of this review – that it may remove poorly liganded iron [2045] and interact with hydroxyl radical directly [2046-2049].

It is notable that appropriate levels of erythropoietin appear not only to be efficacious but to be safe, even in pregnancy [2050-2057]. Erythropoietin may itself be a marker of hypoxia and oxidative stress in pregnancy [447; 2058-2061], consistent with a view that the body is attempting to deal with these problems by creating anti-inflammatory cytokines.

# Hypoxia-inducible factor (HIF)

Although we are mainly not concentrating on genetic regulatory aspects in this review, the HIF [2062; 2063] does deserve some mention, since many of the syndromes described above are accompanied by hpoxia, and this causes levels of the HIF to increase. HIF is a transcription factor that can activate a considerable number of genes, including VEGF [1779; 2062-2066]. In contrast to the constitutive expression of HIF-1$\alpha$, HIF-1$\beta$ protein levels are regulated in response to the cellular oxygen concentration [2067]. The active HIF is the HIF-1$\alpha\beta$ heterodimer [2068]. HIF couples anoxia to innate immunity via the NF-$\kappa$B system [2069]. In particular, HIF effects (via hepcidin) the mobilisation or iron and can cause the expression of inflammatory cytokines such as IL-1,IL-6 and TNF-$\alpha$ [2070-2072] under conditions (hypoxia) where superoxide and peroxide production are likely to be increased, and consequently increases sepsis (in that HIF-knockout mice are resistant to LPS-induced sepsis [2070; 2071]). By contrast, induction of HIF (and the genes that it activates) can effect neuroprotection [2068; 2073]. HIF also appears to have a significant role in placental development, and defective HIF expression may be involved in pre-eclampsia and intra-uterine growth retardation [354; 2062; 2074]. Qutub and colleagues provide useful models [2075; 2076] of HIF activation under a variety of conditions of iron, $O_2$, 2-oxoglutarate and other factors.

## *Autocatalysis, positive feedback and Systems Biology*

What has emerged in recent years is a recognition that the structure of the modules of metabolic and signalling networks, somewhat independent of the individual activities of their components, can have a profound controlling influence on their behaviour (e.g. [1933; 1956; 1957; 2077; 2078]). Classically, negative feedback structures are considered to confer stability, while positive feedbacks tend to have the opposite effect. However, negative feedbacks with delay loops can cause oscillations [1850; 1935] while some kinds of positive feedback loops can confer stability [2078; 2079]. However, there is no doubt that structures in which a damaging agent causes the production of a second damaging agent that itself catalyses the production of the first or a separate damaging agent can exhibit a runaway kind of damage. This is <u>exactly</u> what can happen with iron and superoxide since Fe(n) can be liberated from ferritin by superoxide radicals and then catalyse the production of further hydroxyl radical by increasing the amount of free iron (Fig 6). A similar effect can occur with Fe-S proteins in SOD-deficiency [1030], with the degradation of mitochondrial by radical damage leading to further production of radicals [5; 2080; 2081], and the effects of oxidative stress on iron storage [2082]. This again illustrates the importance of acting at multiple points in a network to control these kinds of damage. Exactly the same is true of the IL-1 and TNF-α systems in which IL-1 or TNF-α (oxidative stress) acting on one cell can effect the secretion of further IL-1/TNF-α that can act on adjacent cells (Fig 8), of the hypoxia-dependent increase in both ROSs and serum iron mediated by hepcidin (Fig 3), the autocatalytic synergy between overfeeding, inflammation and (pre-)diabetes, and of the peroxide/iron pair that are liberated when frataxin is deficient (see above). It is these kinds of synergistic effects and autocatalytic cycles that are the hallmark of the major and progressive effects on human physiology that are seen in these kinds of system.

## Predictive biology

It is often considered (e.g. [2083; 2084]) that a desirable feature of a scientific idea is its ability to make useful predictions, and while this is not in fact a particularly well founded philosophical principle, it probably is of value to set out a couple of 'predictions' that follow from the present analysis. One prominent feature of the above is the primacy of the iron-catalysed production of the damaging hydroxyl radical, and thus a test of the involvement of these kinds of reactions in the various physiological and pathological states to which we allude is the prediction that they should be accompanied by markers of oxidative stress characteristic of reactions of endogenous metabolites with the hydroxyl radical. While it is not that easy to disentangle the complex reactions of ROSs with biomolecules [16; 945], at least the following appear to be a result of reactions involving $OH^{\bullet}$ [2085; 2086]: 8-oxo-2'-deoxyguanosine (oxo$^8$dG) [68; 2087; 2088], 8-oxo-7,8- dihydro-2'-deoxyguanosine [60; 62] and thymine glycol [62] [2085].

Another set of predictions from the systems biology perspective is that combinations [1535-1539; 1562; 2089; 2090] of chemical agents will be far more efficacious in modulating iron-catalysed oxidative stress and its sequelae than will be the use of 'magic bullet' single agents.

Iron-mediated oxidative stress is arguably the cause of much of the inflammation typically observed in biological systems, often further mediated via pro-inflammatory cytokines. Another major prediction that comes from the above then is that molecules that are anti-inflammatory, whether widely recognised as such or not, should have beneficial effects in syndromes for which they have not necessarily been tested. An obvious set of candidates in this regard are to be found among the statins, since it is now clear that they have important anti-inflammatory properties (see above). Thus, there are already indications that as well as their established benefits in cardiovascular disease (e.g. [699; 2091]) they may exert benefit in a huge variety of syndromes [731], including sepsis [732; 1887; 2092-2104], heart failure [2105], pain perception [2106], lupus and related diseases [1163; 2107], diabetes [770; 2108], rheumatoid arthritis [759; 762; 783; 2109-2114], kidney disease [2115-2117], inflammatory skin disease [2118], emphysema [2119], ischemia-reperfusion injury [2120], stroke [757; 765; 2121-2128], traumatic brain injury [2129-2131], neurodegenerative diseases [753-755; 813; 1164; 1886; 2132-2148], neurotoxicity [2149] and cancer [2150-2164].

## Concluding remarks

"Actually, the orgy of fact extraction in which everybody is currently engaged has, like most consumer economies, accumulated a vast debt. This is a debt of theory and some of us are soon going to have an exciting time paying it back – with interest, I hope." [2165].

"If you are not thinking about your experiments on a whole-genome level you are going to be a dinosaur". J. Stamatoyannopoulos, quoted in [2166].

While it is less common for scientists to publish 'negative' results ('there was no effect of some agent on some process'), and there has been a tendency to seek to falsify specific hypotheses rather than to paint a big picture [2167], there is no doubt that the sheer weight of positive evidence can be persuasive in leading one to a view. As Bertrand Russell put it [2168] "When one admits that nothing is certain one must, I think, also admit that some things are much more nearly certain than others." However, the huge volume of scientific activity has in many ways led to a 'balkanisation' of the literature [2169] in which scientists deal with the problem of the deluge of published papers by necessarily ignoring most of them. This is no longer realistic in an age of post-genomics, the internet, Web 2.0 and systems biology, and when we are starting to move to integrative (if distributed) models of organisms (including humans) at a whole organ, genome or whole organism scale [87; 1957; 2170-2176]. With expression profiling atlases becoming increasingly widespread (e.g. [2177; 2178]), and with the ability to exchange models of biochemical networks in a principled manner [2179; 2180] and, when they are marked up appropriately (e.g. [2181; 2182]), to begin to reason about them automatically [87; 2183], we may soon look forward to an era in which we can recognise the commonalities across a variety of different subfields. Thus, therapies derived for one of the inflammatory diseases listed above may well have benefit in some of the others where their underlying 'causes' are the same. In this sense, the combinatorial roles of poorly liganded iron and reactive oxygen species, as set out in the above, appear to be prototypical.

## Acknowledgments

I thank Jon Barasch (Columbia University) for drawing my attention to the role of NGAL in human physiology, and Phil Baker, David Brough and Louise Kenny for many further useful and enjoyable discussions. I thank Katya Tarasova for considerable assistance with literature gathering. Much of my work is supported by the UK BBSRC and EPSRC, for which I am most grateful, and I am also a current recipient of funding from the British Heart Foundation. The funders had no role in study design, data collection and analysis, decision to publish, or preparation of the manuscript. This is a contribution from the Manchester Centre for Integrative Systems Biology (http://www.mcisb.org).

# Legends to figures

Fig 1. An overview of this article, set out in the form of a 'mind map' [34].

Fig 2. Schematic overview of the main elements considered to participate in mammalian iron metabolism.

Fig 3. Some effects of hepcidin, summarizing the fact that hypoxic condition can suppress it and thus lead to hyperferraemia. Since hypoxic conditions can also lead to ROS production the hypoxia-mediated regulation of hepcidin can have especially damaging effects.

Fig 4. Overview of the roles of ischaemia, ROSs, poorly liganded iron and the iron metabolism regulators HGAL and hepicidin in effecting inflammation as a physiological level.

Fig 5. Role of inflammation caused by hydroxyl radical formation in the interactive development of obesity, the metabolic syndrome and diabetes. Intervention at multiple steps is likely to be most beneficial in alleviating this kind of progression.

Fig 6. Catalysis and autocatalysis in the Haber-Weiss and Fenton reactions leading to the production of the hydroxyl radical, including the liberation by superoxide of free iron from ferritin.

Fig 7. Some iron chelators that are in clinical use (left hand side) or that have been proposed.

Fig 8. An overview of cellular signaling using the NF-κB and p38 systems. Note that some of the extracellular effectors that mediate NF-κB activation are themselves produced and secreted as a result of the activation, potentially creating an autocatalytic system.

Fig 9. General view of the role of iron, antioxidants and ROSs in aging and degenerative processes. Some of the decay may be ameliorated by lifestyle and dietary means. Based in part on [1049].

# Tables

**Table 1. Comparison of the main available iron chelators to an ideal chelation drug (modified from [2184])**

| | "Ideal chelator" | Deferoxamine | Deferiprone | Deferasirox |
|---|---|---|---|---|
| Route of administration | Oral | Parenteral, usually subcutaneous or intravenous | Oral | Oral |
| Plasma half-life | Long enough to give constant protection from labile plasma iron | Short (minutes); requires constant delivery | Moderate (< 2 hours). Requires at least 3-times-per-day dosing | Long, 8-16 hours; remains in plasma at 24 h |
| Therapeutic index | High | High at moderate doses in iron-overloaded subjects | Idiosyncratic side effects are most important | Probably high in iron overloaded subjects[*] |
| Molar iron chelating efficiency; charge of iron (III) complex | High, uncharged | High (hexadentate); charged | Low (bidentate); uncharged | Moderate (tridentate); uncharged |
| Important side effects | None or only in iron-depleted subjects | Auditory and retinal toxicity; effects on bones and growth; potential lung toxicity, all at high doses; local skin reactions at infusion sites | Rare but severe agranulocytosis; mild neutropenia; common abdominal discomfort; erosive arthritis | Abdominal discomfort; rash or mild diarrhoea upon initiation of therapy; mild increased creatinine level |
| Ability to chelate intracellular cardiac and other tissue iron in humans | High | Probably lower than deferiprone and deferasirox (it is not clear why) | High in clinical and in *in vitro* studies | Insufficient clinical data available; promising in laboratory studies |

[*]Nephrotoxicity that has been observed in non-iron-loaded animals has been minimal in iron-overloaded humans, but effectiveness is demonstrated only at higher end of tested doses, as discussed in [1527].

cytotoxicity via inhibition of mitochondrial dysfunction and ROS production. *J Neurol Sci* **253,** 53-60.

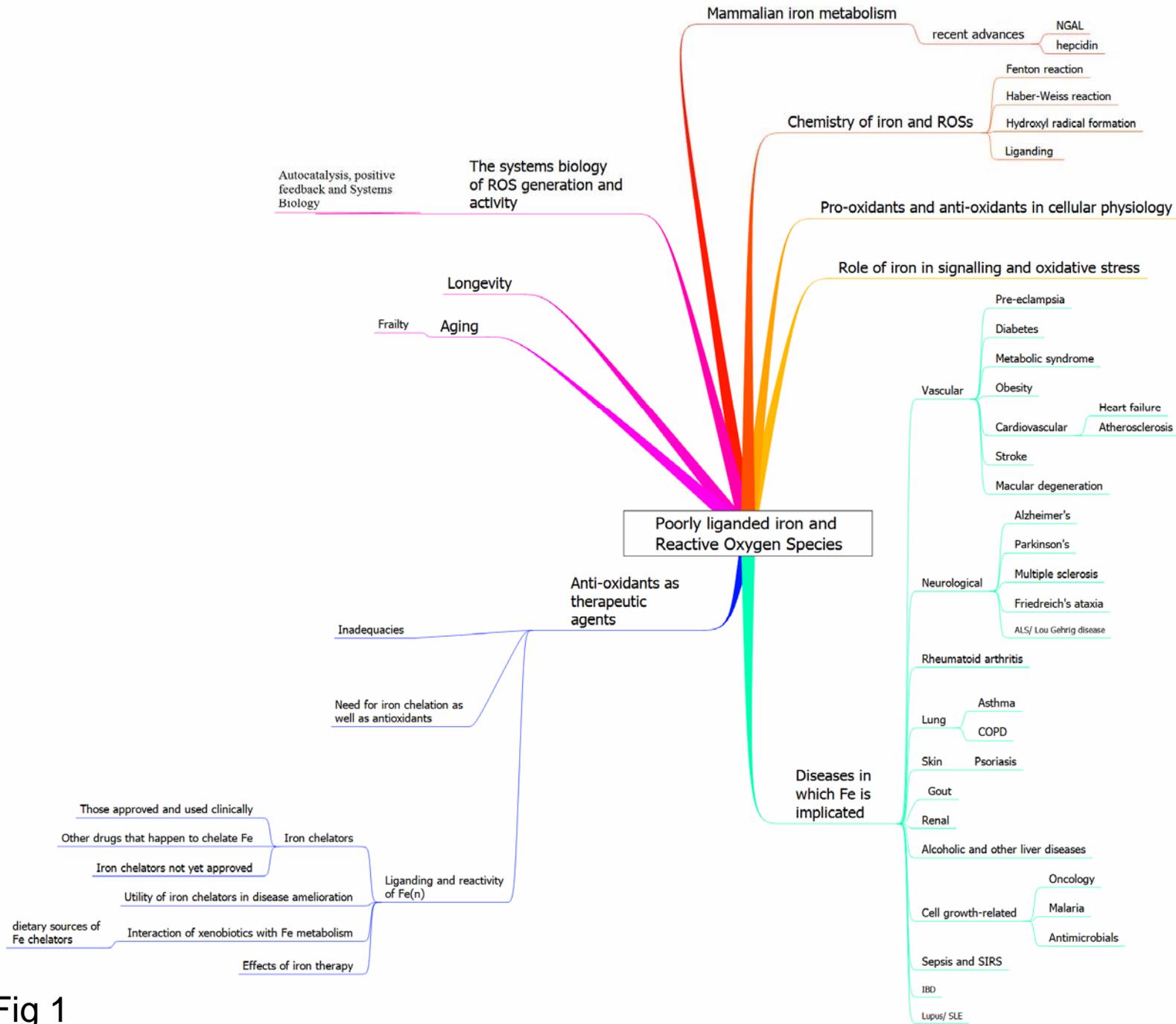

Kell, Fig 1

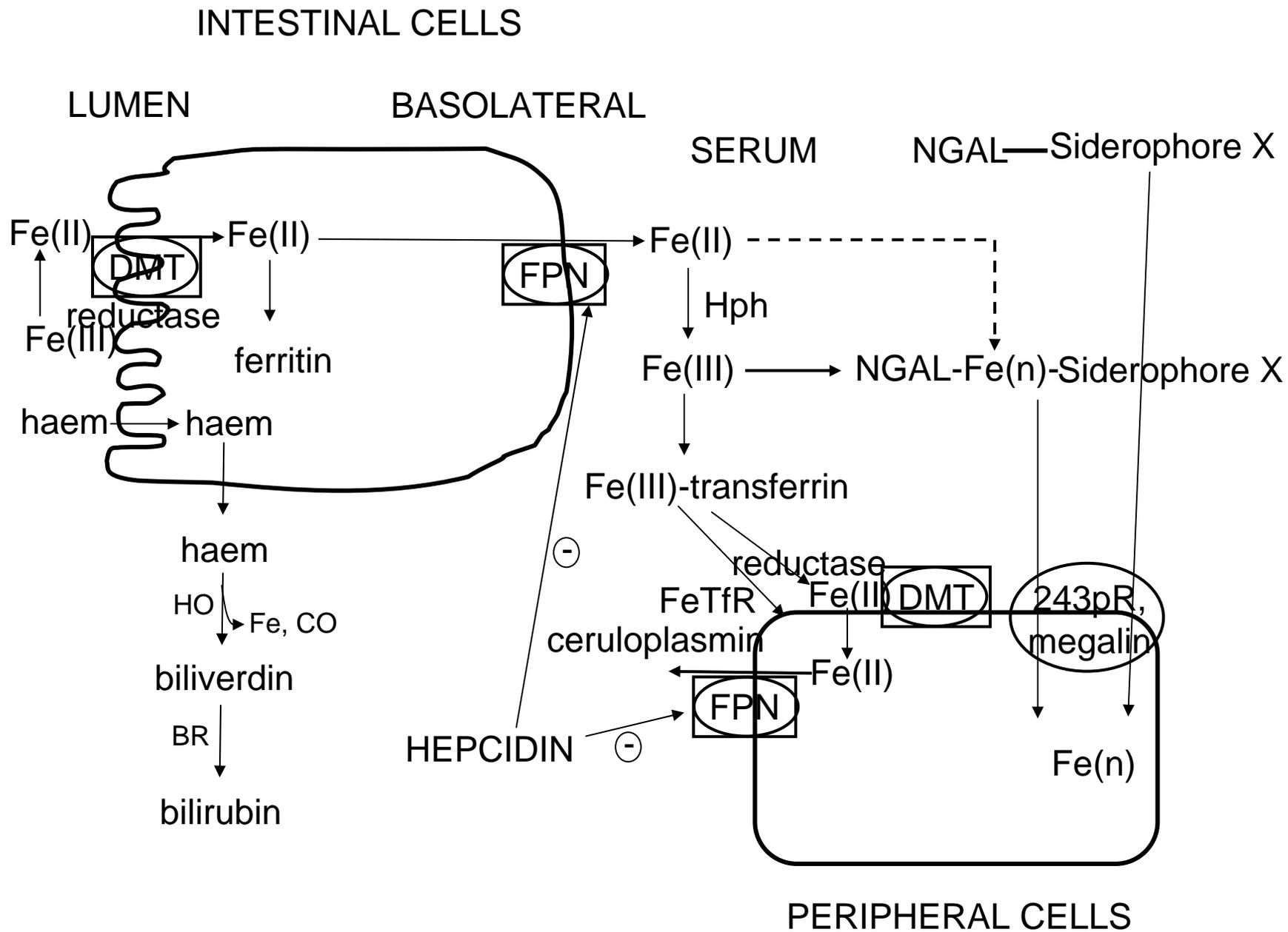

Kell, Fig 2

# Some effects of Hepcidin

- Suppressed by hypoxia
- Raised by high iron stores and/or inflammation
- Hepcidin increase causes internalisation of FPN, with decrease in efflux of Fe(II) from peripheral cells, causing IL-6-mediated hypoferraemia (but thus more Fe(n) to remain intracellular)
- Lowered hepcidin causes hyperferraemia

Kell, Fig 3

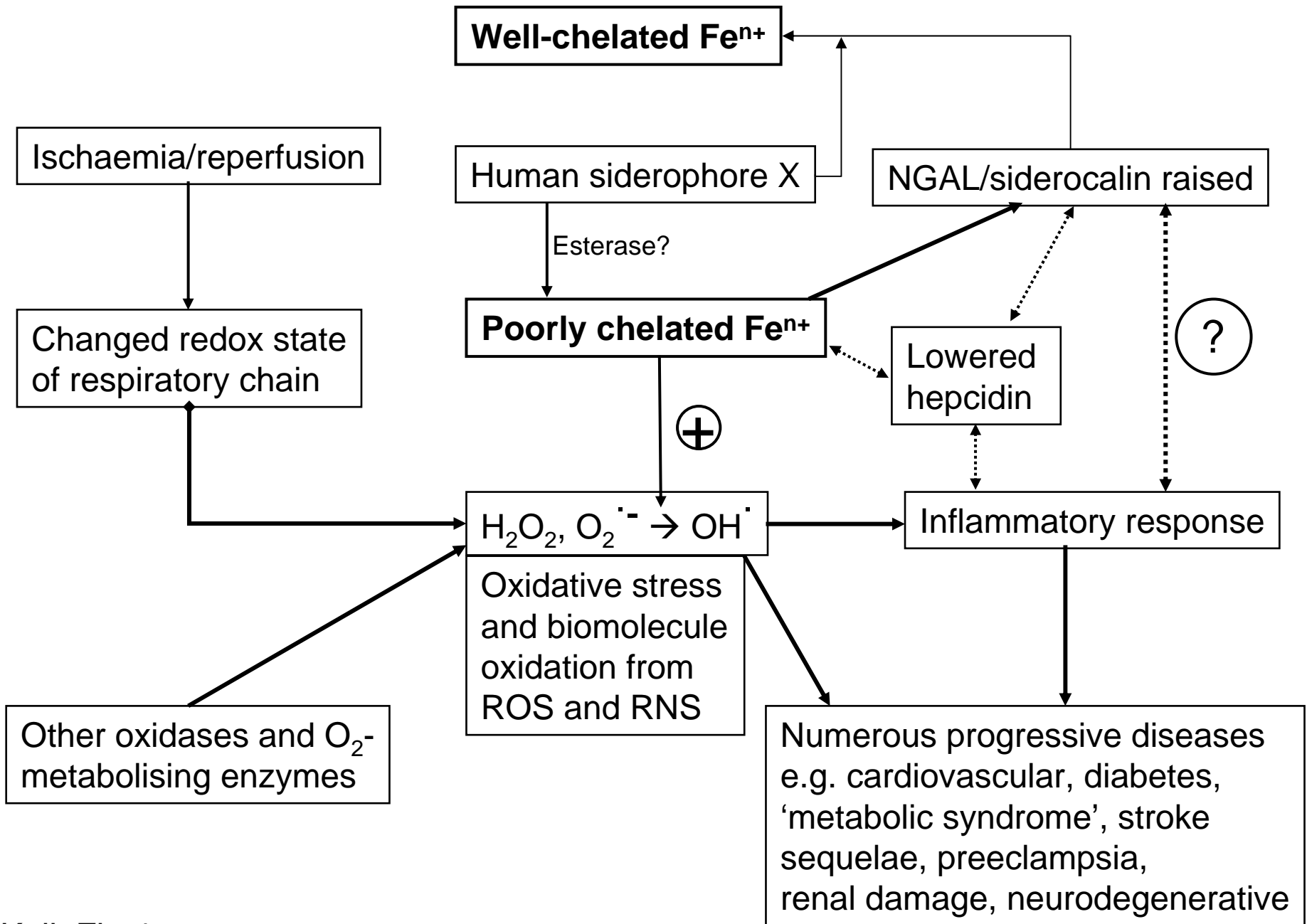

Kell, Fig 4

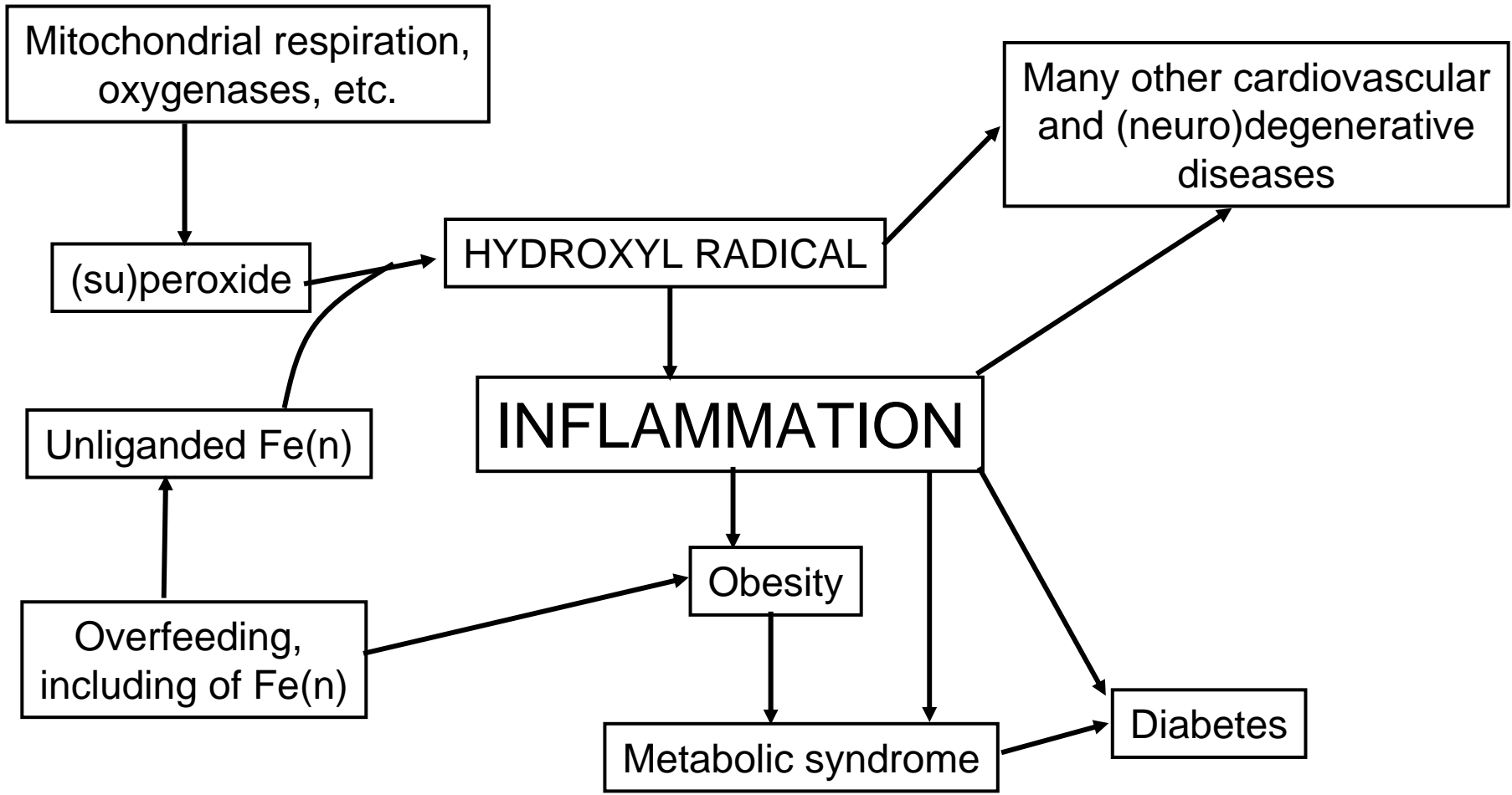

Kell, Fig 5

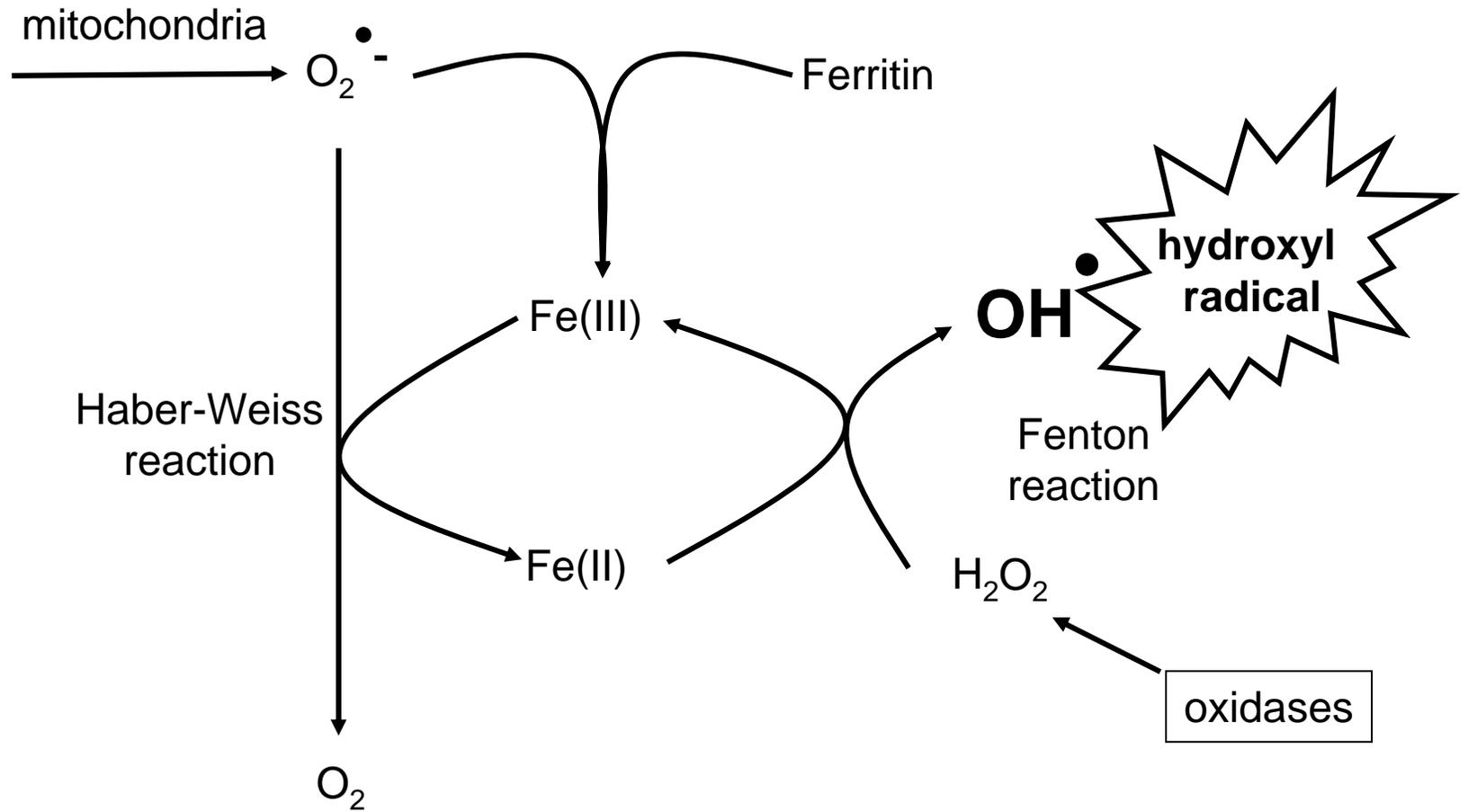

Kell, Fig 6

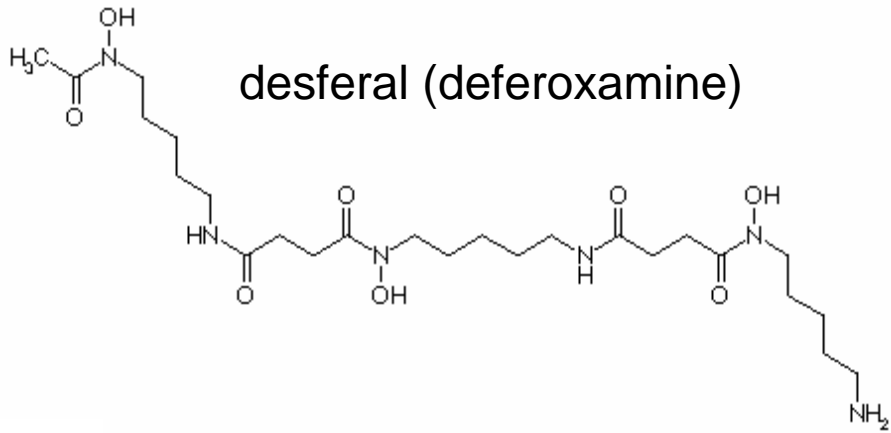

desferal (deferoxamine)

clioquinol 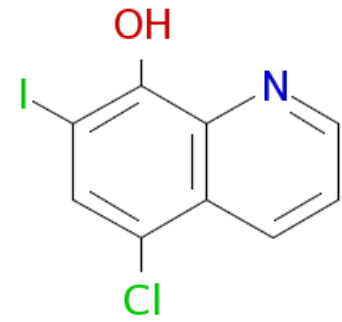

xanthurenic acid

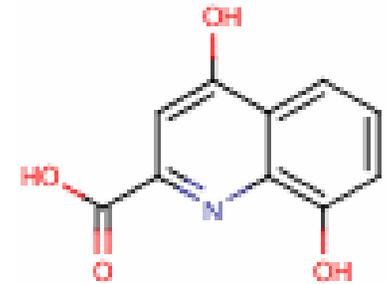

ferriprox (L1 or deferiprone)

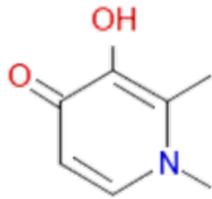

exjade (ICL670 or deferasirox)

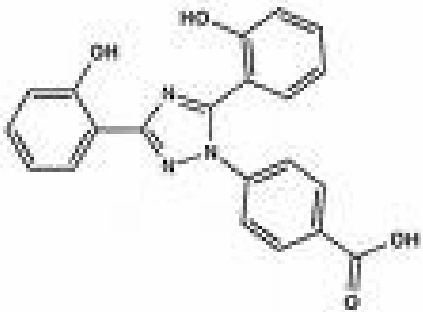

HBED 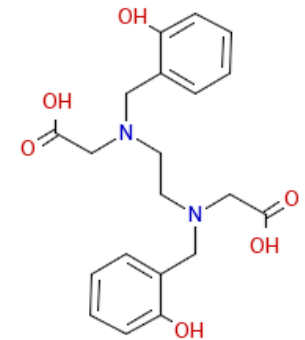

Kell, Fig 7

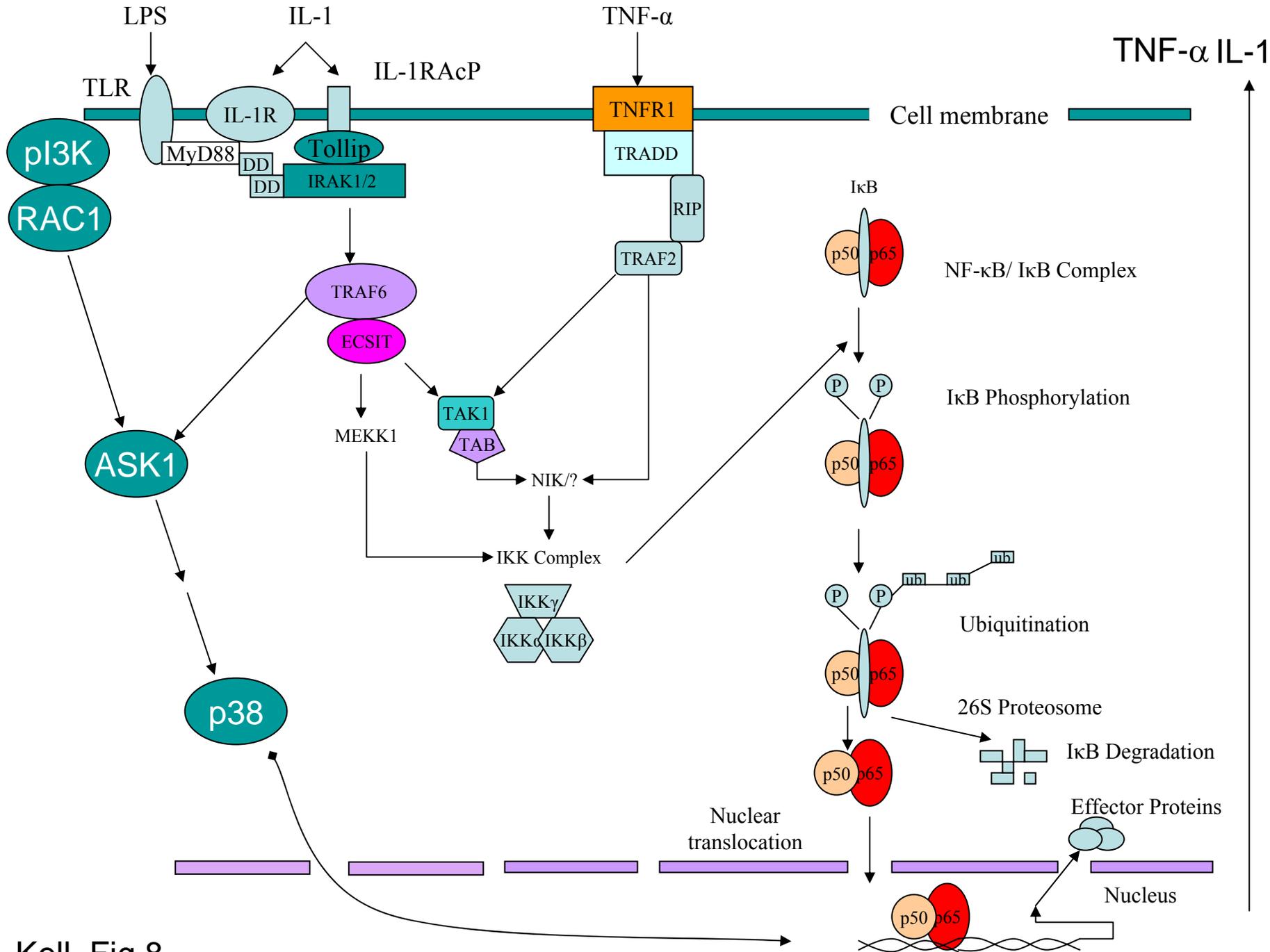

Kell, Fig 8

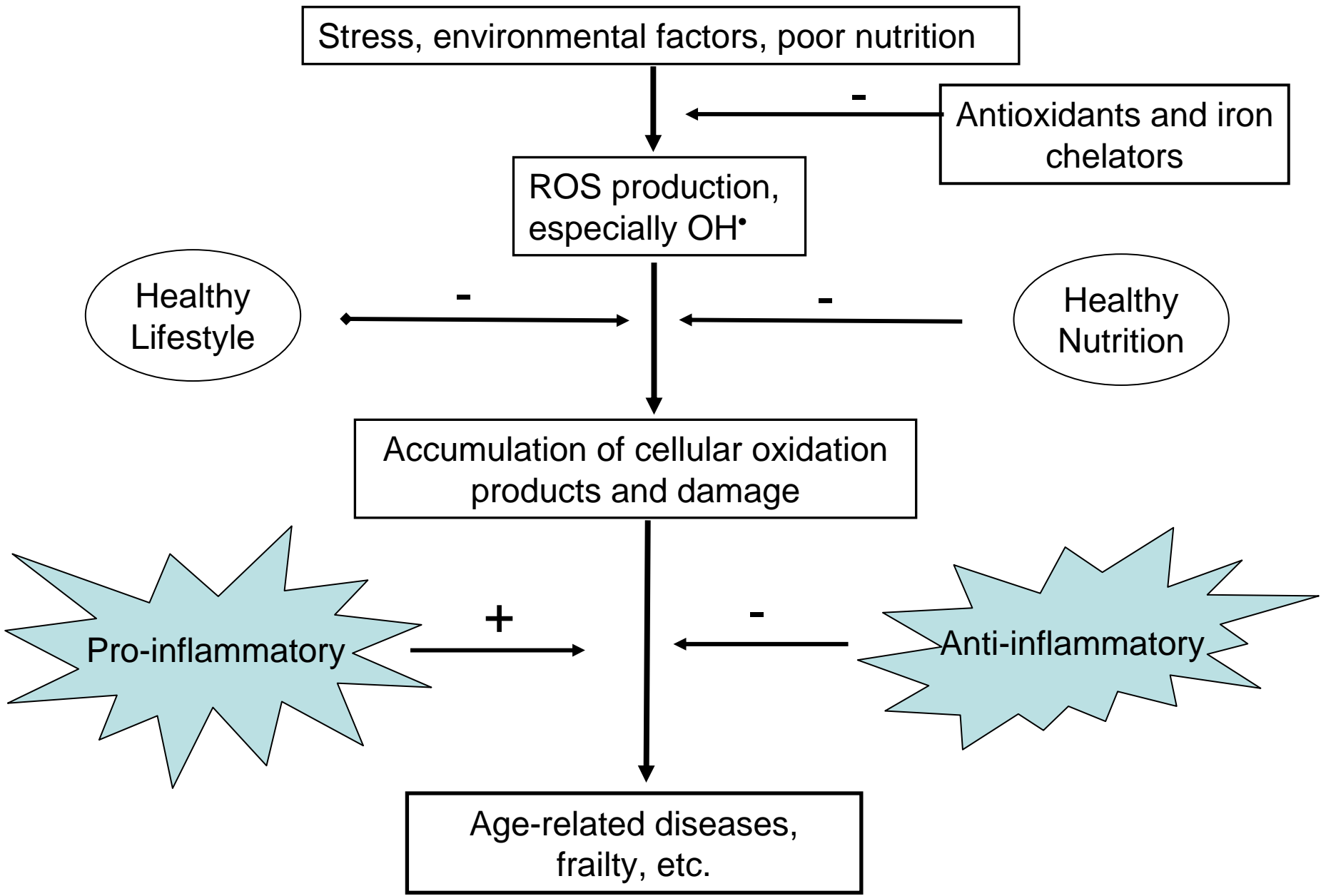

Kell, Fig 9